\documentclass[aps,pra,twocolumn,groupedaddress]{revtex4-1}
\usepackage{amsmath}
\usepackage{graphicx}
\usepackage{amsfonts}
\usepackage{color}
\usepackage{dcolumn}% Align table columns on decimal point
\usepackage{bm}% bold math
\usepackage{multirow}
\usepackage{array}
\usepackage{booktabs}
\usepackage{ctable}
\usepackage{upgreek}
\usepackage{epsfig}
\usepackage{mathrsfs}
\usepackage{amssymb}
\usepackage{amsbsy}
\usepackage{cancel}
\usepackage{pifont}
\usepackage{marginnote}
\usepackage{float}
\makeatletter

\begin{document}

\relax

%\title{Sub-diffusion and non-equilibrium probes of phases in Aubry-Andr{\'e}-Harper Model}
%\title{Super-diffusion to sub-diffusion due to edge effects in Aubry-Andr{\'e}-Harper model}
\title{Anomalous transport in the Aubry-Andr{\'e}-Harper model in  isolated and  open systems}

\date{\today}

\author{Archak Purkayastha}
\affiliation{International centre for theoretical sciences, Tata Institute of Fundamental Research, Bangalore - 560089, India}
\author{Sambuddha Sanyal }
\affiliation{International centre for theoretical sciences, Tata Institute of Fundamental Research, Bangalore - 560089, India}
\author{Abhishek Dhar}
\affiliation{International centre for theoretical sciences, Tata Institute of Fundamental Research, Bangalore - 560089, India}
\author{Manas Kulkarni}
\affiliation{International centre for theoretical sciences, Tata Institute of Fundamental Research, Bangalore - 560089, India}

\begin{abstract}
We study the high temperature transport behavior of the Aubry-Andr{\'e}-Harper (AAH) model, both in the isolated thermodynamic limit and in the open system. At the critical point of the AAH model, we find hints of super-diffusive behavior from the scaling of spread of an initially localized wavepacket. On the other hand, when connected to two baths with different chemical potentials at the two ends, we find that the critical point shows clear sub-diffusive scaling of current with system size. We provide an explanation of this by showing that the current scaling with system-size is entirely governed by the behavior of the single particle eigenfunctions at the boundary sites where baths are attached. We also look at the particle density profile in non-equilibrium steady state of the open system when the two baths are at different chemical potentials. We find that the particle density profile has distinctly different behavior in the delocalized, critical and localized phases of the AAH model.
\end{abstract}

%\maketitle must follow title, authors, abstract, \pacs, and \keywords
\maketitle
\section{Introduction} 

The absence of diffusion in non-interacting systems due to the presence of disorder is referred to as Anderson localization and has been
theoretically studied and experimentally observed in a wide class of systems, e.g for electrons, photons, cold-atoms and sound waves \cite{Mirlin_Evers,Physics_Today_1,Physics_Today_2}. The effect of localization is strongest in one dimension where it is known that a small amount of
disorder localizes all states. If interaction is switched on in such a localized system, a transition from the many-body localized (MBL) phase to delocalized phase can happen. The physics of the system close to the transition is not well understood and has received a lot of attention lately. One of the most interesting results of recent investigations is that close to the transition one has  Griffiths effects leading to slow dynamics and sub-diffusive transport \cite{Griffiths4,Griffiths3,Griffiths2,Griffiths1}. 

An interesting class of models emerge when the `true' disorder is replaced by a quasiperiodic potential. A paradigmatic example of such a system is the so-called Aubry-Andr{\'e}-Harper (AAH) model \cite{aa1,harper}. This is a one-dimensional lattice model of non-interacting particles (bosons or fermions) in an incommensurate potential. For this system one finds a remarkable transition from  all energy eigenstates being localized to all states being extended as one decreases the strength of the potential. This transition is mediated by a critical point \cite{aa1}.  Unlike the MBL transition, this transition occurs in absence of interactions. Also, Griffiths physics is not expected at this critical point because the potential is spatially correlated. Early studies found interesting features at the critical point such as fractal patterns in the spectrum and the eigenstates \cite{fractal1,fractal2,pandit83}. It has also been extensively studied in the mathematical literature \cite{math1,math2}.   The AAH model and its various generalizations have received a lot of interest recently both theoretically \cite{Yang2017,david2017,Naldesi2016,AAH2017,AAH1,AAH2,AAH3,AAH4,AAH5,AAH6,AAH7,AAH8,AAH9,AAH10,AAH11,AAH12,AAH13,AAH14,AAH15,AAH16,AAH17,AAH18,AAH19} and experimentally \cite{expt0,expt1,expt2,expt3,expt4,expt5,expt6}. 

Although studies of wavepacket spreading in the closed system have shown hints of anomalous diffusion (as opposed to normal diffusion) behaviour at the critical point  \cite{wavepacket1,wavepacket2,wavepacket3,wavepacket4}, the exact nature of transport at the critical point has remained an open question. In the context of MBL, it has become most relevant because recent experiments investigating MBL physics are based on AAH model with interactions, rather than a `truly' disordered system \cite{expt0,expt1,expt2,expt3,expt5}.  There has also been a lot of recent interest in disordered interacting systems connected to baths \cite{expt1,MBL_open_1,MBL_open_2,MBL_open_3,MBL_open_4,MBL_open_5,MBL_open_6,MBL_open_7,
MBL_open_8,MBL_open_9,MBL_open_10}. However, there have been only few studies on the open AAH system \cite{AAH6,AAH13}. In particular, there are no results on the non-linear response of the system to external thermal or chemical potential biases. 

In this work, we study the transport properties of the AAH model both in the isolated thermodynamic limit and in the open system. In the isolated thermodynamic limit we look at spread of an initially localized wavepacket and the conductivity calculated by Green-Kubo formalism via numerical exact diagonalization. At the critical point, the integrated current auto-correlation appearing in the Green-Kubo conductivity, seems to saturate to a constant value but with large fluctuations. Correspondingly we find that the second moment for the spread  of the wavepacket goes as $\sim t$, and correctly gives the constant value in the Green-Kubo computation. However, we find that the tails of the wavepacket spreads super-diffusively. As a result, at very long times, the moments show a crossover from diffusive to super-diffusive behavior. This crossover occurs at shorter times for higher moments. A careful quantitative investigation shows that the time scales required to observe this crossover in the second moment is beyond our current computational power. This explains the normal-diffusive-like behavior of Green-Kubo conductivity and suggests that eventually, at extremely long times, the integrated current auto-correlation will diverge. 

Next, we study the open system by connecting to two baths at the two ends. The baths are modelled by quadratic Hamiltonians with infinite degrees of freedom. We calculate the non-equilibrium steady state (NESS) current and particle density profile numerically exactly via the non-equilibrium Green's function (NEGF) approach. At the critical point, we find clear sub-diffusive scaling of current with system size, which is in sharp contrast to the properties of the isolated thermodynamic system discussed above. We provide an explanation of this by showing that the current scaling with system-size is entirely governed by the  behavior of the single particle eigenfunctions at the boundary sites where baths are attached.  We further show that the NESS particle spatial density profile provides a real-space experimentally measurable probe of the localized, critical and de-localized phases.    

In Section.~\ref{model}, we introduce the model, in Section.~\ref{isolated} we discuss the formalism and the results for transport behavior of the isolated system in the thermodynamic limit, in Section.~\ref{open} we discuss the formalism and the results for the open system NESS, in Section.~\ref{conclusions} we give the conclusions.

\section{Model}\label{model} 

The AAH model is given by the Hamiltonian 
\begin{equation}
\label{H_S}
\mathcal{H}_S~=~\sum_{r=1}^{N-1}(\hat{a}_r^{\dagger} \hat{a}_{r +1}+h.c)+\sum_{r=1}^N 2\lambda \cos(2\pi b r+\phi) \hat{a}_r^{\dagger} \hat{a}_r 
\end{equation}
where $b$ is an irrational number and $\phi$ is an arbitrary phase, $\hat{a}_r$ correspond to fermionic (bosonic) annihilation operators defined respectively on $r$-th lattice point of the system of $N$ sites. The hopping parameter has been set to $1$, and this is taken as the energy scale.  When $\lambda<1$, all the energy eigenstates of this model are delocalized and when $\lambda>1$, all energy eigenstates are localized. $\lambda=1$ is the critical point. This holds true for any choice of irrational number $b$ and phase $\phi$. The most popular choice for $b$ is the golden mean $(\sqrt{5}-1)/2$. However, in experiments and numerics all numbers are essentially rational in a strict mathematical sense. The way around is given by the fact that for a system of finite size $N$, if $b$ is taken as a rational number $p/q$ with $q>N$, $b$ remains `effectively irrational' and all the observed physics of AAH model is retained. In  recent experiments \cite{expt0,expt1,expt2}, physics of AAH model has been explored by superimposing a  $532nm$ optical lattice with a $738nm$ one, making  $b=532/738$. For $q<N$, the system becomes delocalized. Even though the choice of $b$ is irrelevant for various interesting universal features of the AAH model, the exact nature of plots depend on $b$. In this work, we have considered the following choices of $b$ : golden mean $(\sqrt{5}-1)/2$, silver mean $\sqrt{2}-1$ and the  rational number $532/738$ used in the experiments in \cite{expt0,expt1,expt2}. Further, we perform an average over the phase $\phi$ by numerically exactly integrating the final results between $0$ and $2\pi$ and dividing by $2\pi$.

\section{Transport in the isolated system in the thermodynamic limit}\label{isolated}
\subsection{Formalism}
We first look at transport properties of the isolated system in the thermodynamic limit. For this we directly calculate particle conductivity of the system using the Green-Kubo formula. For this we define,
\begin{align}
\mathcal{G}(t) = \int_0^{\beta} d\lambda \sum_{p,q=0}^{N-1}\langle \hat{I}_p(-i\lambda)\hat{I}_{q}(t)\rangle / N
\end{align}
where $\hat{I}_p = i (\hat{a}_p^{\dagger} \hat{a}_{p+1} - \hat{a}_{p+1}^{\dagger} \hat{a}_p)$, and $\langle ... \rangle = Tr(e^{-\beta(\mathcal{H}_S - \mu \mathcal{N}_S)}/Z ...)$. $\mathcal{N}_S=\sum_r \hat{a}_r^{\dagger} \hat{a}_r$ is the total number of particles in the system, and $Z=Tr(e^{-\beta(\mathcal{H}_S - \mu \mathcal{N}_S)})$. The conductivity by Green-Kubo formalism is given by 
\begin{align}
\sigma_{GK} = \lim_{\tau\rightarrow \infty} \lim_{N \rightarrow \infty} D_N(\tau)
\label{numlim}
\end{align}
where 
\begin{align}
D_N(\tau)=\int_0^{\tau} \mathcal{G}(t) dt.  
\end{align}
The order of limits in the Green-Kubo conductivity formula is important and cannot be interchanged, and the formula is strictly valid only for infinite system size. But in numerics, one will always have a finite size. To go about numerically calculating Green-Kubo conductivity, one has to look at the behaviour $D_N(\tau)$ for given system size, for times before the finite-size effects become substantial. 

One can show  that the Green-Kubo formula can be related to the spread of correlations.  We start with  the mixing assumption, expected to be valid in the thermodynamic limit. This says that, given two arbitrary operators $Q_1$ and $Q_2$, $\lim_{t\rightarrow \infty} \langle Q_1(t) Q_2(0)\rangle = \lim_{t\rightarrow \infty} \langle Q_1(t) \rangle \langle Q_2 \rangle$. Under this assumption, and time-translation and time-reversal symmetries, the Green-Kubo formula can be simplified to the form,
\begin{align}
\label{sig}
\sigma_{GK} =  \beta \lim_{\tau\rightarrow \infty} \lim_{N \rightarrow \infty} \int_0^\tau dt \sum_{p,q=0}^{N-1} \textrm{Re} \left(\langle \hat{I}_p(t)\hat{I}_{q}(0)\rangle \right) / N
\end{align}
Starting from the continuity equation $\frac{d\hat{n}_p}{dt}=\hat{I}_p - \hat{I}_{p-1} $, where $\hat{n}_p={\hat{a}}_p^\dagger\hat{a}_p$, it can be shown for the infinite size system  that
\begin{align}
\label{continuity_simpf}
\frac{d}{d\tau} \sum_{x=-\infty}^{\infty} x^2 \langle \hat{n}_0(0) \hat{n}_x(\tau) \rangle = 2 \int_0^{\tau}  dt \sum_{x=-\infty}^{\infty}  \langle \hat{I}_0(0) \hat{I}_x(t) \rangle
\end{align}
where we have used time-translation invariance as well as the space translation invariance. The space-translation invariance is not present for our particular model in Eq.~\ref{H_S}, but  this is restored for quantities averaged over $\phi$. Now, using translation invariance, it follows from Eqs.~(\ref{sig},\ref{continuity_simpf}) that
\begin{align}
\label{sigma_simpf}
\sigma_{GK}= \lim_{\tau\rightarrow \infty} \frac{\beta}{2}\frac{d}{d\tau} \textrm{Re} \left( \sum_{x=-\infty}^\infty x^2 \langle \hat{n}_0(0) \hat{n}_x(\tau) \rangle \right)~. 
\end{align}
Note that here the $N\rightarrow \infty$ limit has already been taken before while using Eq.~\ref{continuity_simpf}.
For normal diffusive transport, 
\begin{align}
\label{m_2nn}
m_2^{nn}(\tau)= \textrm{Re} \left(\sum_{x=-\infty}^\infty x^2 \langle \hat{n}_0(0) \hat{n}_x(\tau) \rangle \right)= 2 D \tau,
\end{align}
for large $\tau$.
Thus, the Green-Kubo conductivity for normal diffusive transport is given by
\begin{align}
\label{sigma_analyt}
\sigma_{GK} =\lim_{\tau\rightarrow \infty} \lim_{N \rightarrow \infty} D_N(\tau) = {\beta D}~.
\end{align}
Hence, for normal diffusive transport, we expect that, for large enough $N$, $D_N(\tau)$ will converge to this value as $\tau$ increases, before finite-size effects become substantial. 

In general, $m_2^{nn}(t)\sim t^{2\tilde{\beta}}$. As seen above $\tilde{\beta}=0.5$, for normal diffusive transport. For ballistic transport, $\tilde{\beta}=1$. If $0.5<\tilde{\beta}<1$, transport is super-diffusive. For both super-diffusive and ballistic transports, as seen from Eqs.~\ref{sigma_simpf},\ref{m_2nn}, the $\sigma_{GK}$ diverges. If $0<\tilde{\beta}<0.5$, the transport is sub-diffusive, while for a localized system, $\tilde{\beta}=0$. In both these cases, $\sigma_{GK}$ is zero. All cases other than the normal diffusive transport are broadly classified as anomalous transport.

This brings us to the study of $\langle \hat{n}_0(0) \hat{n}_x(\tau) \rangle$. In a very recent paper \cite{david2017}, spread of a similar quantity was studied to classify transport behaviour for the AAH model with interactions. 
Since our system is non-interacting, $\langle \hat{n}_x(t)\hat{n}_{0}(0)\rangle$ can be written down in terms of the single particle eigenfunctions. We have $\hat{c}_\ell(t) = \sum_{p=1}^{N} G(\ell,t \mid p, 0) \hat{c}_p(0)$, where $G(\ell,t \mid p, 0)$ is the single particle Green's function for the closed system. Let $\Phi_{\alpha , \ell}$ be the $\ell$th component of the single-particle eigenvector corresponding to the single-particle energy eigenvalue $\epsilon_\alpha$. Thus $\hat{c}_\ell =\sum_{\alpha=1}^{N} \Phi_{\alpha , \ell} \tilde{ \hat{c}}_\alpha$ where $\tilde{ \hat{c}}_\alpha$ are the annihilation operators in the eigenbasis. Here $\alpha$ is eigenstate index and $\ell$ is the site index.  Then $G(\ell,t \mid p, 0) = \sum_{\alpha=1}^N e^{-i\epsilon_\alpha t} \Phi_{\alpha , \ell} \Phi_{\alpha , p}$. In terms of these, we have,
\begin{align}
\label{nxt}
&C(x,t)=\langle \hat{n}_x(t)\hat{n}_{0}(0)\rangle - \langle \hat{n}_x \rangle \langle \hat{n}_0 \rangle \nonumber \\
&=\Big[ \sum_{\ell,p=1}^N G^*(x+{N}/{2},t \mid \ell, 0) G(x+{N}/{2},t \mid p, 0) \nonumber \\
&\langle \hat{c}_\ell^{\dagger} (0) \hat{c}_{N/2}(0)\rangle \langle \hat{c}_p (0) \hat{c}_{N/2}^{\dagger}(0)\rangle \Big] \nonumber \\
&=\sum_{\alpha,\nu=1}^{N} \Big [ \Phi_{\alpha , x+N/2} \Phi_{\nu , x+N/2} \Phi_{\alpha , N/2} \Phi_{\nu , N/2} \nonumber \\
&\times e^{i(\epsilon_\alpha -\epsilon_\nu)t} n_F(\epsilon_\alpha)(1-n_F(\epsilon_\nu) \Big ],
\end{align}
where $n_F(\omega)=\left[e^{\beta(\omega-\mu)}+ 1\right]^{-1}$ is the fermi distribution function. 

\begin{figure}
\includegraphics[width=\columnwidth]{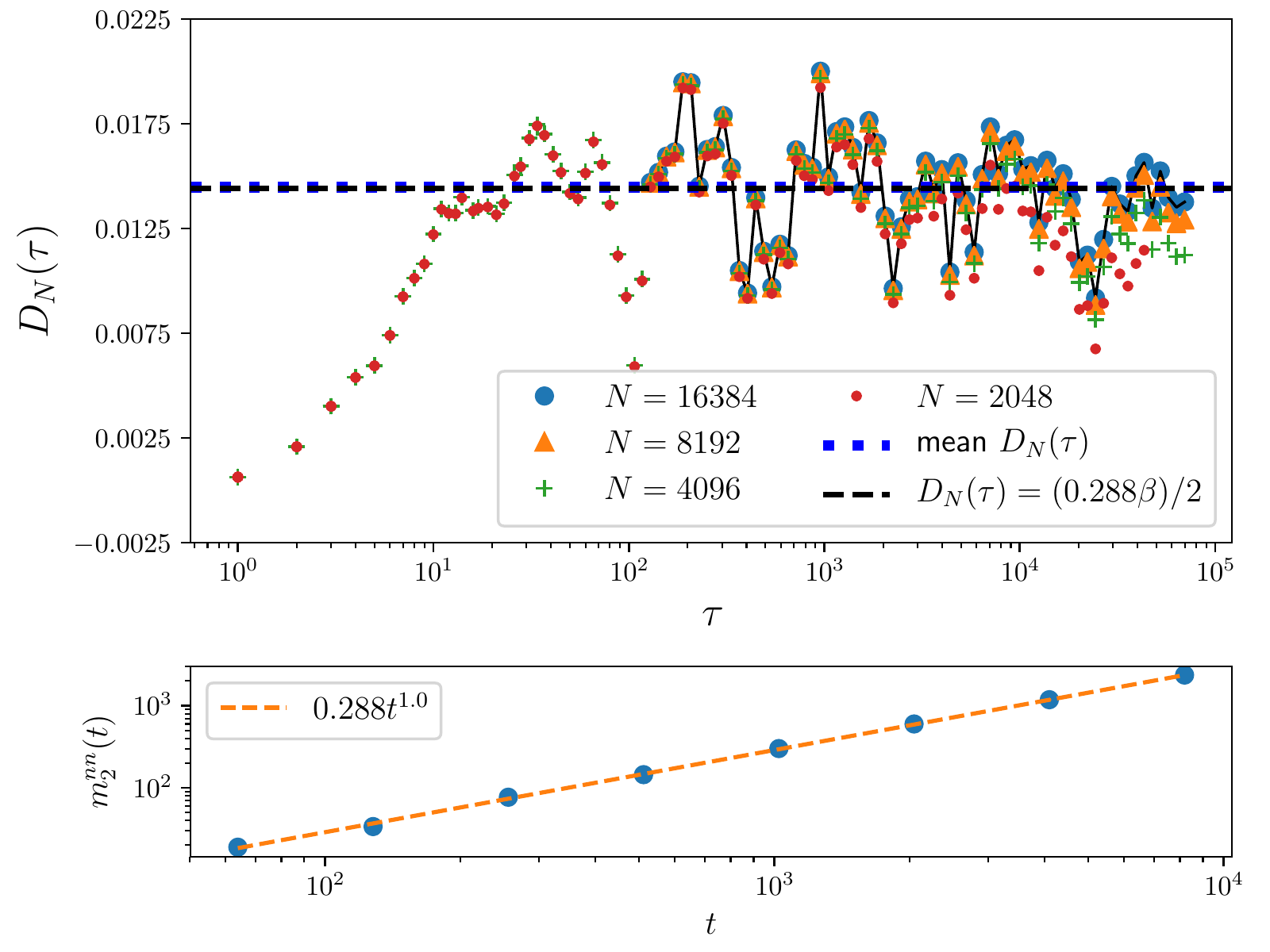}
\caption{(color online) \textbf{Isolated thermodynamic limit}:  Top panel shows plot of  $D_N(\tau)$ as a function of $\tau$ at the critical point $\lambda=1$ for different system sizes.  $D_N(\tau)$  initially increases with $\tau$ and then shows large fluctuations about a constant mean value. This mean value is quite precisely given by the analytical high temperature approximation result $D_N(\tau)\simeq (D\beta)/2$. $D$ is obtained by time scaling of $m_2^{nn}$ (Eq.~\ref{m_2nn}) shown in bottom panel. $D=0.288/2$. For $N=8192, 16384$, only the large time results have been calculated. The black continuous line is guide-to-eye joining data points for $N=16384$. The mean $D_N(\tau)$ is calculated from the data points for $N=16384$.   Parameters: $\beta=0.1$,  $\mu=1$, $b=\sqrt{2}-1$. }
\label{kubofig}
\end{figure}

\begin{figure*}
\includegraphics[width=\linewidth]{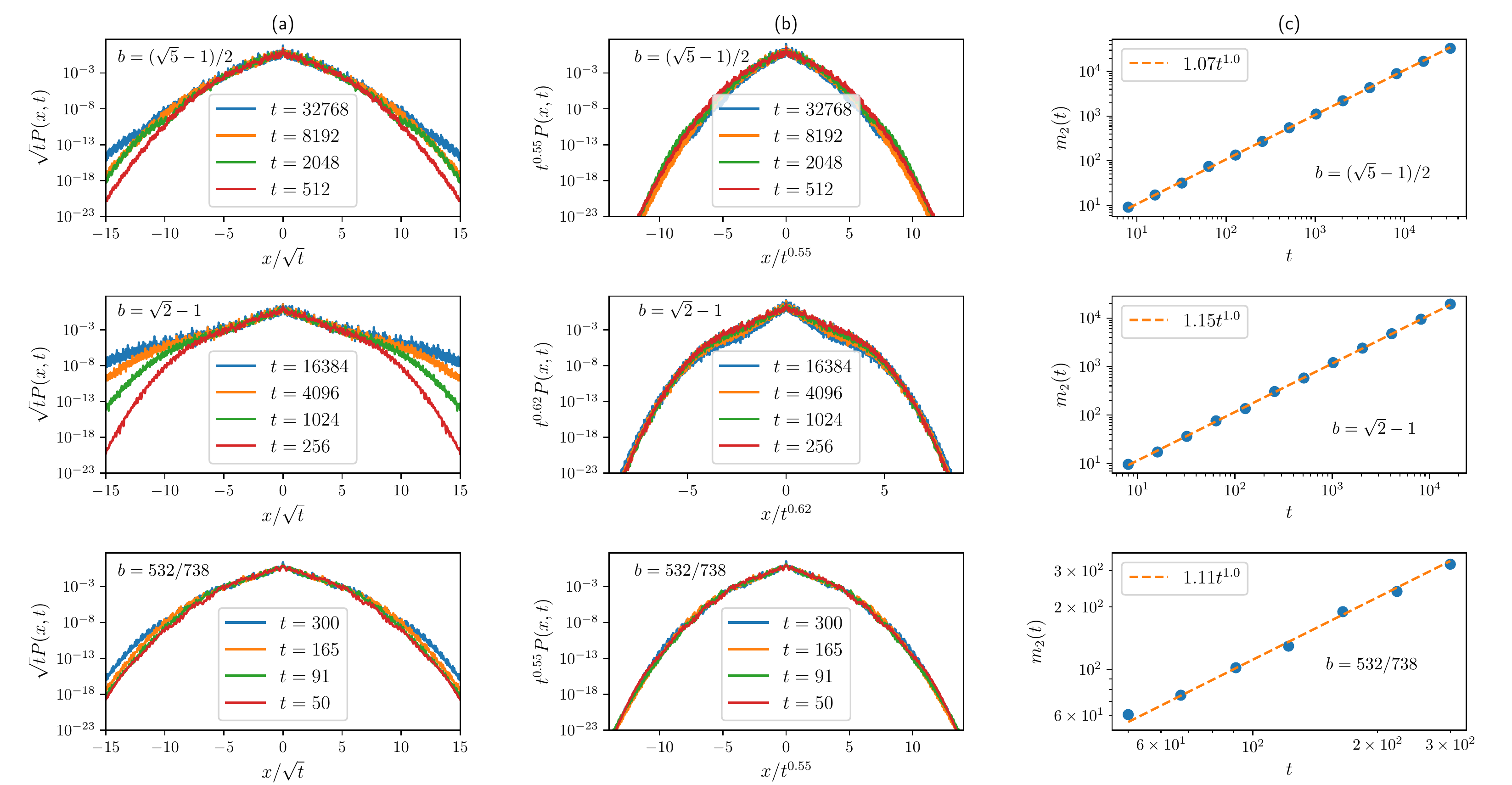}
\caption{(color online) \textbf{Isolated thermodynamic limit}: \textbf{(a)} The full distributions $P(x,t)=~\mid\psi(x,t) \mid ^2$, scaled assuming normal diffusive behaviour.  Here $x=r-N/2$.  $P(x,t)$ scales as $P(x,t)\simeq (1/\sqrt{t}) f_1({x}/{\sqrt{t}})$ over a considerable region in the bulk but the scaling function $f_1(z)$ is clearly not Gaussian and also the tails do not collapse. \textbf{(b)} $P(x,t)=~\mid\psi(x,t) \mid ^2$ scaled to collapse the tails of the distribution. The tails show a super-diffusive scaling $P(x,t)\simeq (1/t^{\tilde{\beta}}) f_2({x}/{t^{\tilde{\beta}}})$ with $\tilde{\beta}>0.5$. However, the value of $\tilde{\beta}$ depends on the choice of $b$. \textbf{(c)} The scaling of second moment $m_2(t)$ of $P(x,t)$ with $t$ for various values of $b$. $m_2(t)\sim t$.  $N=8192$ for $b=(\sqrt{5}-1)/2, \sqrt{2}-1$. $N=700$ for $b=532/738$.  }
\label{Pxtplot}
\end{figure*}

Another related quantity that has been studied previously \cite{wavepacket1,wavepacket2,wavepacket3,wavepacket4} for the AAH model is the spread of a wavepacket. Let the wavepacket $\psi_r(t)$ be initially localized at the site $N/2$ of the lattice. 
It evolves according to the Schroedinger equation 
\begin{align}
i{\partial \psi_r}/{\partial t}=\psi_{r+1}(t)+\psi_{r-1}(t)+2\lambda \cos(2\pi b r+\phi)\psi_{r}(t). 
\end{align}
We  define $x=r-N/2$ and look at the probability $P(x,t)=\mid\psi_x (t)\mid^2$, and its moments 
\begin{align}
m_{2p}(t)=\sum_{x=-N/2}^{N/2-1} (x-\langle x \rangle)^{2p} P(x,t),
\end{align}
where $\langle x \rangle =\sum_{x=-N/2}^{N/2-1} xP(x,t)$ is the mean. In terms of single-particle wave functions $P(x,t)$ is given by,
\begin{align}
\label{Pxt}
&P(x,t) = \mid G(x+{N}/{2},t \mid {N}/{2}, 0) \mid^2 \nonumber\\
&=  \sum_{\alpha,\nu=1}^{N} \Phi_{\alpha , x+N/2} \Phi_{\nu , x+N/2} \Phi_{\alpha , N/2} \Phi_{\nu , N/2} e^{i(\epsilon_\alpha -\epsilon_\nu)t} .
\end{align}
This is different from $C(x,t)$ (see Eq.~\ref{nxt}) by only the factor  $n_F(\epsilon_\alpha)(1-n_F(\epsilon_\nu)$ inside the summation. At high temperatures, $n_F(\omega)\sim 1/2$. Thus, at high temperatures, 
\begin{align}
\label{Cxt_Pxt}
\beta\rightarrow 0, ~~ C(x,t)\rightarrow \frac{P(x,t)}{4} 
\end{align}
We are interested in the high temperature transport. So the scaling properties of $C(x,t)$ and $P(x,t)$, and hence of $m_2^{nn}(t)$ and $m_2(t)$ will be same.

For normal diffusive spreading, $P(x,t)$ has a Gaussian form,  $P(x,t)=e^{-\frac{x^2}{16Dt}}/{\sqrt{16\pi D t}}$, this $D$ being being the same as defined in Eq.~\ref{m_2nn}. In general, if $P(x,t)$ (and thereby $C(x,t)$) scales as 
\begin{align}
\label{Pxt_scaling}
P(x,t)\sim ({1}/{t^{\tilde{\beta}}})f({x}/{t^{\tilde{\beta}}}),
\end{align}
then $[m_{2n}(t)]^{1/n}\sim t^{2\tilde{\beta}}$. The connection to different regimes of transport as discussed before is immediate. However, there may occur cases where, $\tilde{\beta}=0.5$ but, $P(x,t)$ is not Gaussian. Such `non-Gaussian but diffusive' transport has been reported in many classical systems \cite{nongauss_diff0,nongauss_diff1,nongauss_diff2,nongauss_diff3,nongauss_diff4,nongauss_diff5,
nongauss_diff6,nongauss_diff7}. This is also considered as anomalous transport. Further, there may even occur cases where $P(x,t)$ does not follow a particular scaling form. As shown in Eq.~\ref{sigma_simpf}, even then, the conductivity of the system depends only the time scaling of second moment, and  classification of transport based on that is possible.

\begin{figure*}
\includegraphics[width=\linewidth]{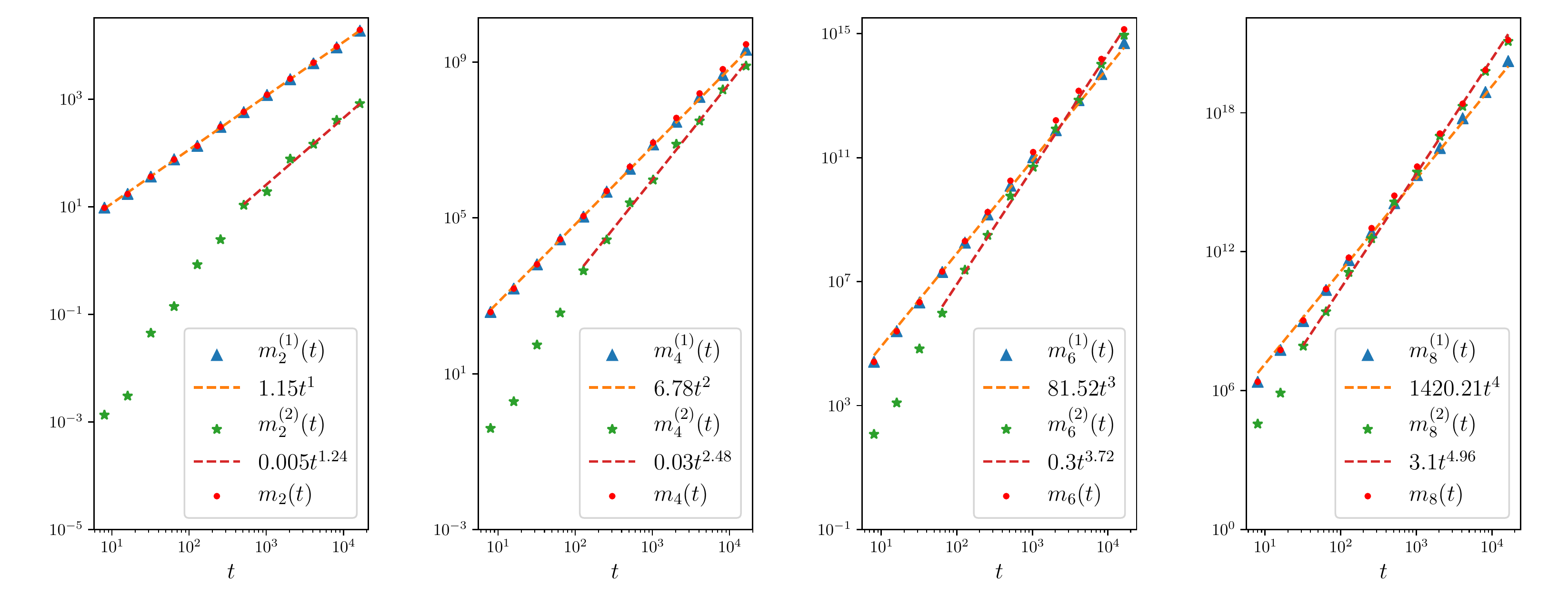}
\caption{(color online) \textbf{Isolated thermodynamic limit}: Plots of $m_{2p}^{(1)}(t)$, $m_{2p}^{(2)}(t)$ and $m_{2p}(t)$ (see Eq.~\ref{m21_m22}) with time for $b=\sqrt{2}-1$. $z_0=5$. The dashed lines are fits for $m_{2p}^{(1)}(t)$ and $m_{2p}^{(2)}(t)$.  $m_{2p}^{(1)}(t)\sim t^p$, whereas, $m_{2p}^{(2)}(t)\sim t^{0.61p}$ for large $t$, as expected from tail scaling of $P(x,t)$. The crossover of $m_{2p}$ scaling from diffusive to super-diffusive is seen clearly for $m_8(t)$ and $m_6(t)$. From the scaling fits, we see that for $m_2(t)$ this crossover will occur for time $t\gg 10^{10}$. $N=8192$.}
\label{m2468}
\end{figure*}

\subsection{Results}
With the above understanding, we numerically investigate transport behavior of the AAH model in the isolated thermodynamic limit via exact diagonalization. All our results are given up to times before finite-size effects become substantial.  We are primarily interested in the transport properties of the AAH model at the critical point ($\lambda=1$). Fig. \ref{kubofig} shows plot of $D_N(\tau)$ with $\tau$ for different system sizes, at the critical point for $b=\sqrt{2}-1$. We see that with increasing $N$, $D_N(\tau)$ converges to a curve which initially increases and then shows large fluctuations about a constant mean value. Consistently, $m_2^{nn}(t)\sim 0.288 t$, and the mean value is precisely given by $\beta D$ (see Eqs.~\ref{m_2nn},\ref{sigma_analyt}). This seems to suggest that $\sigma_{GK}$ is finite in the thermodynamic limit, which is akin to a normal `diffusive' system. However, the fluctuations do not decrease on averaging over $\phi$, and may indicate deviation from normal diffusive transport.

To investigate the nature of transport at the critical point more closely, we now look at the scaling of $P(x,t)$. This is shown in Fig.~\ref{Pxtplot} for various choices of $b$. It is clear that although $m_2(t)\sim t$, $P(x,t)$ is non-Gaussian, and does not obey a single scaling form. The bulk of $P(x,t)$ has the scaling form of Eq.~\ref{Pxt_scaling}, with $\tilde{\beta}=0.5$ for all choices of $b$. However, the tails of $P(x,t)$ do not collapse under the same scaling. This deviation from bulk scaling is most clearly seen for $b=\sqrt{2}-1$. To collapse the tails of $P(x,t)$, one needs a super-diffusive scaling. Thus we find
\begin{align}
\label{Pxt_form}
P(x,t)=
\begin{cases}
\left({1}/{\sqrt{t}} \right)f_1({x}/{\sqrt{t}}) & \forall~\mid x \mid \leq z_0 \sqrt{t} \\
\left({1}/{t^{\tilde{\beta}_2}}\right)f_2({x}/{t^{\tilde{\beta}_2}}),~~\tilde{\beta}_2>0.5 & \forall~ \mid x \mid > z_0 \sqrt{t}
\end{cases}
\end{align}
where $z_0$ and $\tilde{\beta}_2$ depend on the choice of $b$. Note that $z_0$ is independent of time. 
The super-diffusive scaling exponent $\tilde{\beta}_2$ is non-universal and depends on the choice of $b$. For $b=(\sqrt{5}-1)/2$ and for $b=532/738$,   $\tilde{\beta}_2 \sim 0.55$, for $b=\sqrt{2}-1$, $\tilde{\beta}_2\sim 0.62$. 

Note that for $b=(\sqrt{5}-1)/2$ and $b=532/738$, from Fig.~\ref{Pxtplot}(b) it may seem that the super-diffusive scaling of $P(x,t)$ holds everywhere. This is because the super-diffusive exponent $0.55$ is quite close to $0.5$. However, a closer inspection reveals that this is not the case, and the bulk indeed has a diffusive-like scaling. This is clear from the fact that in all cases $m_2(t)$ in Fig.~\ref{Pxtplot}(c) shows the diffusive behavior, $m_2(t)\sim t$, and not the super-diffusive behavior. Also note that, for $b=\sqrt{2}-1$,  $m_2(t)\sim 1.15 t \simeq 4m_{2}^{nn}(t)$, consistent with Eq.~\ref{Cxt_Pxt}.

Now, let us see if the behavior of tails of $P(x,t)$ can affect the scaling of the moments at extremely long times. To check this, we write $m_{2p}(t)$ as
\begin{align}
\label{m21_m22}
m_{2p}(t) &=2\sum_{x=0}^{x\leq z_0 \sqrt{t}} x^{2p} P(x,t) + 2\sum_{x>z_0 \sqrt{t}}^\infty x^{2p} P(x,t) \nonumber \\
& \simeq 2 \int_{0}^{z_0 \sqrt{t}}  x^{2p} P(x,t)dx + 2\int_{z_0 \sqrt{t}}^\infty  x^{2p} P(x,t)dx \nonumber \\
&\equiv m_{2p}^{(1)}(t) + m_{2p}^{(2)}(t)
\end{align}
where $m_{2p}^{(1)}(t)$ is the contribution to moment from the diffusive part, while $m_{2p}^{(2)}(t)$ is the contribution to moment from the tails. Here we have used the fact that $\langle x\rangle=0$, and $P(x,t)$ is an even function of $x$. Now, changing variables to $z_1=x/\sqrt{t}$ and $z_2=x/t^{\tilde{\beta}_2}$, and using
Eq.~\ref{Pxt_form}, we have
\begin{align}
m_{2p}(t) &\simeq 2 t^p \int_{0}^{z_0} z_1^{2p} f_1(z_1) dz_1 \nonumber \\
& + 2t^{2p\tilde{\beta}_2}\int_{z_0 \sqrt{t}/t^{\tilde{\beta}_2}}^\infty z_2^{2p} f_2(z_2) dz_2 \nonumber \\
&= 2 t^{2p\tilde{\beta}_2} \left( t^{p(1-2\tilde{\beta}_2)} A_p + F_p(z_0 t^{0.5(1-2\tilde{\beta}_2)}) \right)
\end{align}
where $A_p=\int_{0}^{z_0} z_1^{2p} f_1(z_1) dz_1$ and $F_p(\tau)=\int_{\tau}^\infty z_2^{2p} f_2(z_2) dz_2$.
Note that $A_p$ is independent of time while $F_p$ is a function of time. So $m_{2p}^{(1)}(t)\sim t^p$, whereas $m_{2p}^{(2)}(t)\sim t^{2p\tilde{\beta}_2}$ only asymptotically. Since $\tilde{\beta}_2>0.5$, we have,
\begin{align}
m_{2p}(t) \sim 2 t^{2p\tilde{\beta}_2} F_p(0),~~ t\rightarrow \infty
\end{align}
Thus, the extreme long time behavior of the moments should be super-diffusive. Hence, there will be a crossover in time scaling of moments from diffusive to super-diffusive. The approach to the super-diffusive scaling is faster for higher moments. Let us check this quantitatively for $b=\sqrt{2}-1$, which is the case where  $\tilde{\beta}_2\sim 0.62$ differs most significantly from the value $0.5$. The value of $z_0$ can be read off from Fig.~\ref{Pxtplot} as $z_0\sim 6$. Fig.~\ref{m2468} shows the plots of $m_{2p}(t)$, $m_{2p}^{(1)}(t)$, $m_{2p}^{(2)}(t)$ for $p=1,2,3,4$. The first thing to note is that the approach to the form $m_{2p}^{(2)}(t)\sim t^{2p\tilde{\beta}_2}$ is faster for higher moments. Secondly, as expected, the crossover to super-diffusive scaling of $m_{2p}(t)$ also occurs faster for higher moments. For $m_{8}(t)$ and $m_6(t)$, this crossover is clearly seen from our data. On the other hand, for $m_4(t)$ and $m_2(t)$, the crossover occurs later than times accessible in our numerics. From the scaling-fits, it is possible to quantitatively extract the time scales at which the super-diffusive crossover will be seen in the $m_2(t)$ scaling. We find that the super-diffusive scaling of $m_2(t)$ will start showing for $t\gg t^*\sim 10^{10}$. To directly investigate such long time behavior without having finite-size effects, one needs systems of size $N \gg (t^*)^{0.62}\sim 10^7$. Exact numerical analysis of such system sizes is definitely beyond our current computational power. This explains the normal-diffusive-like behavior of Green-Kubo conductivity up to times and system-sizes within our numerical reach, and suggests that at even longer times, the super-diffusive behavior will show up.

Therefore, we find \textit{hints of super-diffusive behavior at the critical point of the AAH model in the isolated thermodynamic limit} from the tail scaling of $P(x,t)$ and the time scaling of higher moments of $P(x,t)$. However direct numerical observation of this super-diffusive behavior from $m_2(t)$ scaling or from Green-Kubo conductivity is beyond our current numerical reach. Within our numerical reach, $m_2(t)$ scales diffusively, and Green-Kubo conductivity also shows normal-diffusive-like behavior.

\begin{figure}
\includegraphics[width=\columnwidth]{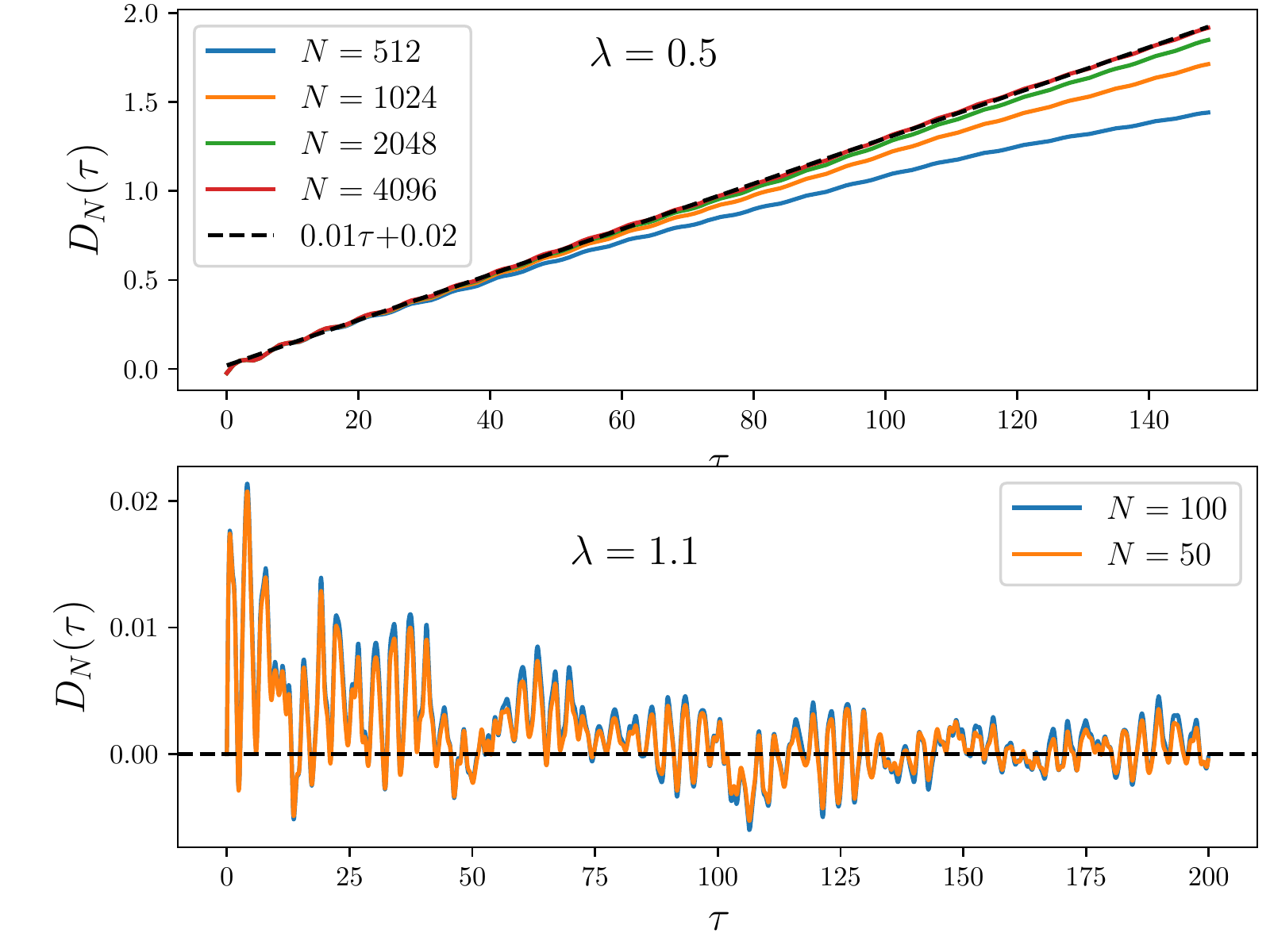}
\caption{(color online) \textbf{Isolated thermodynamic limit}:  Plot showing  $D_N(\tau)$ as a function of $\tau$ for delocalized (top panel) and localized (bottom panel) cases for different system sizes. For delocalized case, $D_N(\tau)$ increases linearly with $\tau$ before finite size effects come into play. For localized case, $D_N(\tau)$ decays to zero and is independent of $N$ for $N\gg$localization length$={1}/{\log(\lambda)}\simeq 10$. Parameters: $\beta=0.1$,  $\mu=1.$, $b=\sqrt{2}-1$. }
\label{Dnt_loc_deloc}
\end{figure}

Away from the critical point, the behavior is exactly as expected. Plots of $D_N(\tau)$ for delocalized and localized regimes are shown in Fig.~\ref{Dnt_loc_deloc} for $b=\sqrt{2}-1$. In the delocalized regime ($\lambda<1$), $D_N(\tau)$ increases linearly with $\tau$ before finite-size effects become significant. Finite size effects start showing after times of $O(N)$. Thus, to numerically take the correct limit (in Eq. \ref{numlim}) for a given system size $N$ one needs to look at $\tau \sim N$. This correctly gives the ballistic conductivity scaling with system size, $\sigma\sim N$. It is also trivial to check $m_2(t)\sim t^2$. In the localized regime ($\lambda<1$), for system sizes much greater than the localization length (given by $1/\log(\lambda)$ \cite{aa1}), the thermodynamic limit is reached and $D_N(\tau)$ becomes independent of $N$. We see $D_N(\tau)$ decays to zero as a function of $\tau$ for such cases, thus giving zero conductivity. Obviously, because all eigenstates are localized, $m_2(t)\sim t^0$ consistently.

We will show below that, when the system is connected to baths, the transport behavior at the critical point of AAH model completely changes.

\section{Transport in the open system}\label{open}
\subsection{Formalism}
Having investigated the transport properties of the isolated AAH model, we now look at transport properties of the open AAH model, i.e, when the AAH system is connected to baths. For this, we couple the  system Hamiltonian $\mathcal{H}_S$ (Eq.~\ref{H_S}) bilinearly with two baths at two ends. The baths are modelled by non-interacting Hamiltonians with infinite degrees of freedom.   
The full Hamiltonian of the system+bath reads as $\mathcal{H}=\mathcal{H}_S+ \mathcal{H}_B+\mathcal{H}_{SB}$, where
\begin{align}
&\hat{\mathcal{H}}_B~=~ \hat{\mathcal{H}}_B^{(1)}+\hat{\mathcal{H}}_B^{(N)}, \nonumber \\ &\hat{\mathcal{H}}_B^{(p)}~=\sum_{s} \Omega_{p s} \hat{B}_{p s}^{\dagger} \hat{B}_{p s}, p=1,N, \nonumber \\
&\hat{\mathcal{H}}_{SB} =  \sum_{s} (\tilde{\kappa}_{p s} \hat{B}_{p s}^{\dagger} \hat{a}_p + h.c.). 
\end{align}
Here $\hat{B}_{s}^{(p)}$ is the annihilation operator of the $s$th mode of the of the bath attached to $p$th site of the system. The baths are connected at the $1$st and the $N$th sites of the system. Here we consider the case where all operators are fermionic. However, since we will be looking at high temperature behavior, the all operators bosonic case will give identical results. We assume that, initially, each of the two baths is at thermal equilibrium at its own temperature and chemical potential. In this paper, we present results for the case when the two baths are at the same temperature but have different chemical potentials thereby having a voltage bias. So, we introduce the bath Fermi distributions 
\begin{align}
n^{(p)}_F(\omega)=\left[e^{\beta(\omega-\mu_p)}+ 1\right]^{-1},~~ p= 1,N. 
\end{align}
But, again, in the high temperature regime, our results remain valid for the case of both thermal and chemical potential biases. Let us also introduce the bath spectral functions
$J_p(\omega)=2\pi \sum_s  \mid \tilde{\kappa}_{p s} \mid^2 \delta(\omega - \Omega_{p s})$.
We assume the two bath spectral functions to be identical $J_1(\omega)=J_N(\omega)=J(\omega)$.

\begin{figure*}
\includegraphics[width=\linewidth]{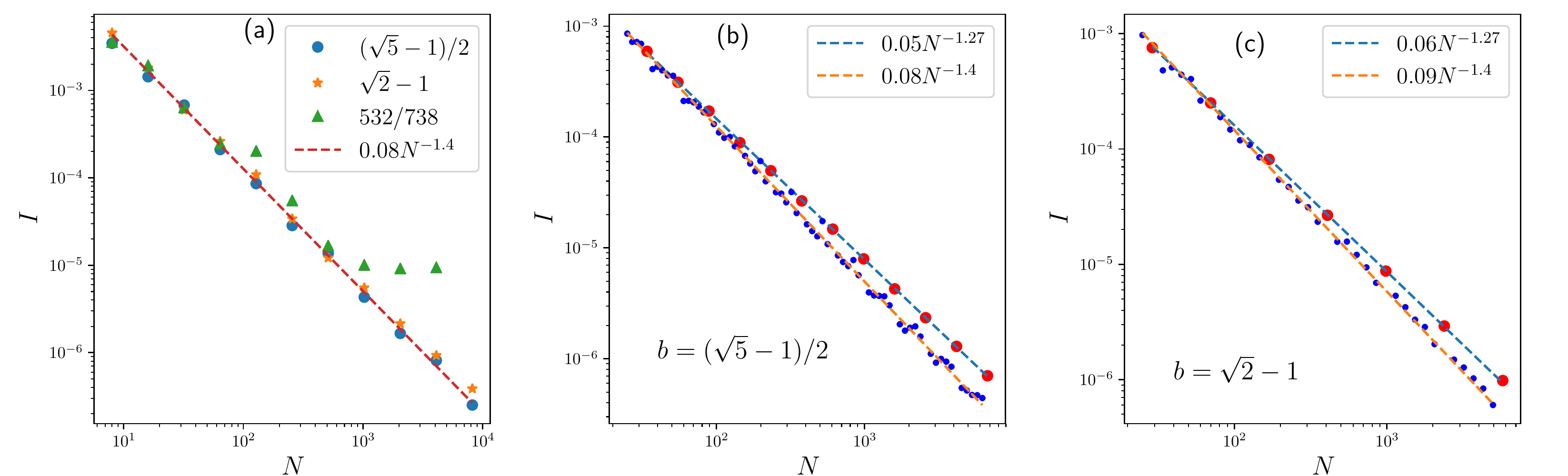}
\caption{(color online) \textbf{Open system}: \textbf{(a)} Scaling of current $I$ with system size $N$ for various values of $b$. $I\sim N^{-1.4\pm 0.05}$. For $b=532/738$, current scaling shows ballistic behavior, $I\sim N^0$ for $N\gg 738$, as expected. Here the system sizes taken are powers of $2$. \textbf{(b)} Scaling of current with system size for $b=(\sqrt{5}-1)/2$ with much closely taken points. This reveals that $I\sim N^{-1.27\pm0.01}$ for $N=$Fibonacci numbers (red circles), whereas  $I\sim N^{-1.4\pm0.05}$ for system sizes away from Fibonacci numbers. \textbf{(c)} Scaling of current with system size for $b=\sqrt{2}-1$ with much closely taken points. This reveals that $I\sim N^{-1.27\pm0.01}$ for $N=$Pell numbers (red circles), whereas  $I\sim N^{-1.4\pm0.05}$ for system sizes away from Pell numbers. Parameters : $\beta=0.1$, $\mu_1=3, \mu_2=-3$, $\gamma=1$, $t_B=3$.}
\label{curr_scaling_1}
\end{figure*}

We are interested in the non-equilibrium steady state (NESS) of this set-up. The NESS properties of this set-up can be exactly calculated via non-equilibrium Green's function (NEGF) formalism. The system Hamiltonian can be written as $\mathcal{H}_S=c^\dagger \mathbf{H}_S c$, with $c$ being the column vector with the $j$th element $c_j=\hat{a}_j$ and $c^\dagger$ is the transpose conjugate. Let $\mathbf{\mathcal{G}}(\omega) = \mathbf{M}^{-1}(\omega)$ be the NEGF of the set-up. $\mathbf{M}(\omega)$ is given by the $N\times N$ matrix $\mathbf{M}(\omega)~=~\left[ \omega\mathbf{I}- \mathbf{H}_S - \mathbf{\Sigma}^{(1)} (\omega)-\mathbf{\Sigma}^{(N)} (\omega)\right]$, where $\mathbf{\Sigma}^{(1)} (\omega)$, $\mathbf{\Sigma}^{(N)} (\omega)$ are   bath self energy matrices with the only non-zero elements given by  $\mathbf{\Sigma}^{(p)}_{pp}(\omega) = -\mathcal{P}\int_{-2t_B}^{2t_B} \frac{d\omega^\prime J(\omega^\prime)}{2\pi(\omega^\prime-\omega)}-\frac{i}{2}J(\omega),$ $p=1,N$, where $\mathcal{P}$ denotes principal value. The NESS quantities of our interest will be the (particle) current $I$ and the occupation of the $r$th site $\langle \hat{n}_r \rangle$. These are given by
\begin{align}
\label{NESS_formulae}
&I=~\int~\frac{d\omega}{2\pi}T(\omega)[~n^{(1)}_F(\omega)-n^{(N)}_F(\omega)], \nonumber \\
&T(\omega)=\frac{J^2(\omega)}{\mid det(\mathbf{M}(\omega))\mid^2}, \nonumber \\
&\langle \hat{n}_r \rangle =~\int \frac{d\omega}{2\pi} \Big[\mid \mathcal{G}_{r1}(\omega) \mid^2 n^{(1)}_F(\omega) + \mid \mathcal{G}_{rN}(\omega) \mid^2 n^{(N)}_F(\omega)\Big],
\end{align}
where $T(\omega)$ is the transmission function. In linear response regime, the expression for (particle) conductance $G$ is given by
\begin{align}
\label{conductance}
G = \beta \int~\frac{d\omega}{2\pi}T(\omega)n_F(\omega)[1-n_F(\omega)]
\end{align}
Note that all information about the explicit model of bath is now in $J(\omega)$.   Different non-interacting baths correspond to different choices of $J(\omega)$. For concreteness, in the following we choose 
\begin{align}
\label{J_def}
J(\omega) = \frac{2\gamma^2}{t_B}\sqrt{1-\left(\frac{\omega}{2t_B}\right)^2},
\end{align}
which can the explicitly derived if baths are modelled via semi-infinte tight-binding chains with hopping parameter $t_B$ and bilinearly connected to the system at the one end via system-bath coupling ${\gamma}$ \cite{ap1}. If all operators were bosonic, only the Fermi distributions in Eq.~\ref{NESS_formulae} would be replaced by the corresponding Bose distribution.

\begin{figure*}
\includegraphics[width=\linewidth]{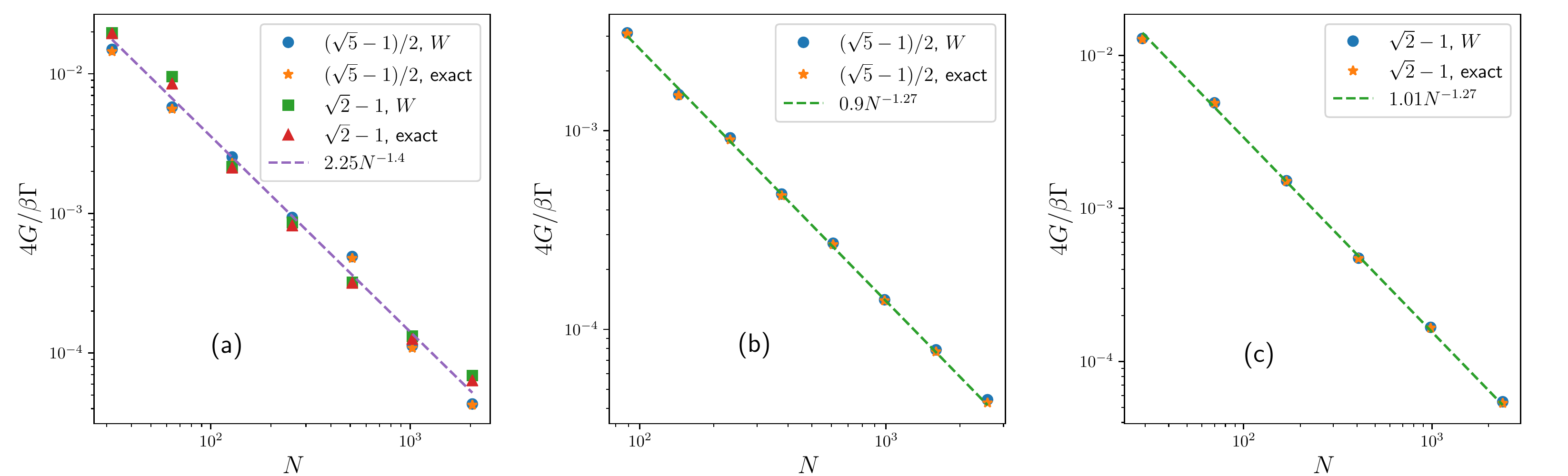}
\caption{(color online) \textbf{Open system}: \textbf{(a)} Scaling of conductance $G$ with system size $N$ for various values of $b$ for weak system-bath coupling ($\gamma=1$, $t_B=200$) and very high temperature ($\beta=0.01$). Exact numerical results are obtained via Eq.~\ref{conductance}, and that is compared with approximate analytical result $W$ (see Eq.~\ref{W}). There is near perfect overlap of exact results with $W$, and $G\sim N^{-1.4\pm0.05}$. Here the system sizes taken are powers of $2$. \textbf{(b)}, \textbf{(c)}   The different scaling of $G$ with system size equal to Fibonacci and Pell numbers for golden mean and silver mean cases, $G\sim N^{-1.27\pm0.01}$. }
\label{W_fig}
\end{figure*}

The main characterization of transport in the open system is via the system size scaling of current $I$. At high temperatures, the current $I$ and the conductance $G$ scale the same way. At very high temperatures, $\beta\rightarrow 0$, $I\simeq  (\mu_1-\mu_2) G \simeq (\mu_1-\mu_2)(\beta/4)\int~\frac{d\omega}{2\pi}T(\omega)$. The conductivity in the thermodynamic limit as obtained from the open system approach is given by $\sigma_{open}= \lim_{N\rightarrow \infty} N G \simeq \lim_{N\rightarrow \infty} N~I/(\mu_1-\mu_2)$. For a diffusive system, conductivity is finite, so $I\sim N^{-1}$. For ballistic transport, current is independent of system size, so $I\sim N^{0}$. If $I\sim N^{-\alpha}$, with $0<\alpha<1$, transport is super-diffusive.  In both super-diffusive and ballistic cases, the $\sigma_{open}$ diverges.  If $I\sim N^{-\alpha}$, with $\alpha>1$, transport is sub-diffusive while  for a localized system, $I\sim e^{-N}$ and in these cases, the $\sigma_{open}$ vanishes. 

The fundamental difference between $\sigma_{open}$ and $\sigma_{GK}$ is the following. In calculating $\sigma_{GK}$, as given in Eq.~\ref{numlim}, the thermodynamic limit $N\rightarrow\infty$ is taken before taking $t\rightarrow\infty$ limit. As a consequence, the system can be considered really isolated and there is no effect of any bath. On the other hand, in calculating $\sigma_{open}$, the $t\rightarrow\infty$ limit is taken first so that the NESS is reached, and then the $N\rightarrow\infty$ limit taken. A detailed and rigorous discussion regarding this is given in \cite{Archak2017}. Physically, in the open system approach, there is the effect of an boundary between the system and the bath, while in the Green-Kubo approach, because of taking the thermodynamic limit first, there is no boundary. As we will see below, the occurrence of the boundary drastically changes the transport properties of the open AAH model at the the critical point. We will also see that the NESS particle density profile has very different behavior is the three phases of the AAH model.

\subsection{Results}

\subsubsection{Current scaling with system size}
The current scaling with system size at the critical point for various choices of $b$ is shown in Fig.~\ref{curr_scaling_1}(a). Here system sizes were taken as powers of $2$. It is immediately clear that the scaling is sub-diffusive with $I\sim N^{-1.4\pm0.05}$. It is also interesting to note that for $b=532/738$ the current becomes independent of $N$ for $N \gtrsim 738$, which is the signature of the delocalized phase. This is consistent with our previous discussion that $532/738$ remains `effectively irrational' only for $N \lesssim 738$. 

In Fig.~\ref{curr_scaling_1}(b), (c), we investigate the current scaling with system size more closely for golden mean and silver mean cases. We see that for golden (silver) mean, the current scaling with system size is different for system sizes equal to Fibonacci (Pell) numbers, where $I\sim N^{-1.27\pm 0.01}$. Away from these special system sizes, the current scaling is approximately $I\sim N^{-1.4\pm0.05}$. An interesting observation follows from noting that any irrational number has an infinite  continued fraction representation which, on truncation, gives a rational approximation of the irrational number. We conjecture that at special system sizes equal to the denominators of the rational approximations, the current deviates from the generic behaviour and has a different scaling. These special system sizes are the Fibonacci (Pell) numbers for the golden (silver) mean.

Thus, we see that the \textit{transport in the open critical AAH model is sub-diffusive}. This is drastically different from what we found in the isolated thermodynamic limit, where we found hints of super-diffusive behavior. We now investigate the origin of the sub-diffusive behavior. To do this, we first take the $t_B\rightarrow$ large limit, so that, the system-bath coupling becomes weak  and the bath spectral functions become almost constant (see Eq.~\ref{J_def}). Note that in Fig.~\ref{curr_scaling_1}, the system-bath coupling was not weak. In the weak system-bath coupling limit, it is possible to express the steady state expressions, involving non-equilibrium Greens functions, directly in terms of the eigenstates and eigenvalues of the isolated system \cite{dhar2017,amir2018}. Using the formalism in Ref.~\cite{ap1}, it can be shown \cite{archak2018} that  for sufficiently small system-bath coupling, the conductance is given by
\begin{align}
\label{W}
G &\simeq \frac{\Gamma\beta}{4}W,~~\beta\rightarrow \textrm{small},~t_B \rightarrow \textrm{large} \nonumber \\
W&=\sum_{\alpha=1}^{N}\frac{\Phi_{\alpha,1}^2\Phi_{\alpha,N}^2}{\Phi_{\alpha,1}^2+\Phi_{\alpha,N}^2}
\end{align}
where $\Gamma=(2\gamma^2)/t_B$ and we have also taken the small $\beta$ limit so that $n_F(\omega)\simeq 1/2$. Thus, the system-size scaling of $G$ in this limit is given by the system size scaling of $W$. Note that $W$ only depends on the absolute values of the single-particle eigenvectors of the system at sites where the baths are attached, namely the first and the last sites. If the system size scaling of $G$ in this limit is similar to that in Fig.~\ref{curr_scaling_1}, which is not guaranteed a priori, we will know that the sub-diffusive scaling is because of the system size scaling of $W$. 

The system size scaling of conductance calculated in this limit ($\gamma=1,t_B=200,\beta=0.01$) by exact numerical integration, Eq.~\ref{conductance}, and by Eq.~\ref{W} is given in Fig.~\ref{W_fig}. There is near perfect overlap of the two results. Note that exact numerical calculation using Eq.~\ref{conductance} is more difficult in this regime, because of the nearly singular behavior of the integrand at system eigenenergies. In Fig.~\ref{W_fig}(a), the scaling is shown for system-sizes in powers of $2$. The scaling is not as good as that seen in the strong system-bath coupling case, but it is approximately the same, $G\sim N^{-1.4\pm0.05}$. Fig.~\ref{W_fig}(b) (Fig.~\ref{W_fig}(c)) shows the scaling for golden (silver) mean case when system sizes are equal to Fibonacci (Pell) numbers. Here there is an almost perfect scaling of $G\sim N^{-1.27\pm0.01}$ as before. Thus, indeed, the  sub-diffusive scaling of current and conductance with system-size is directly related to the  sub-diffusive scaling of $W$ with system-size.

\begin{figure}
\includegraphics[width=\columnwidth]{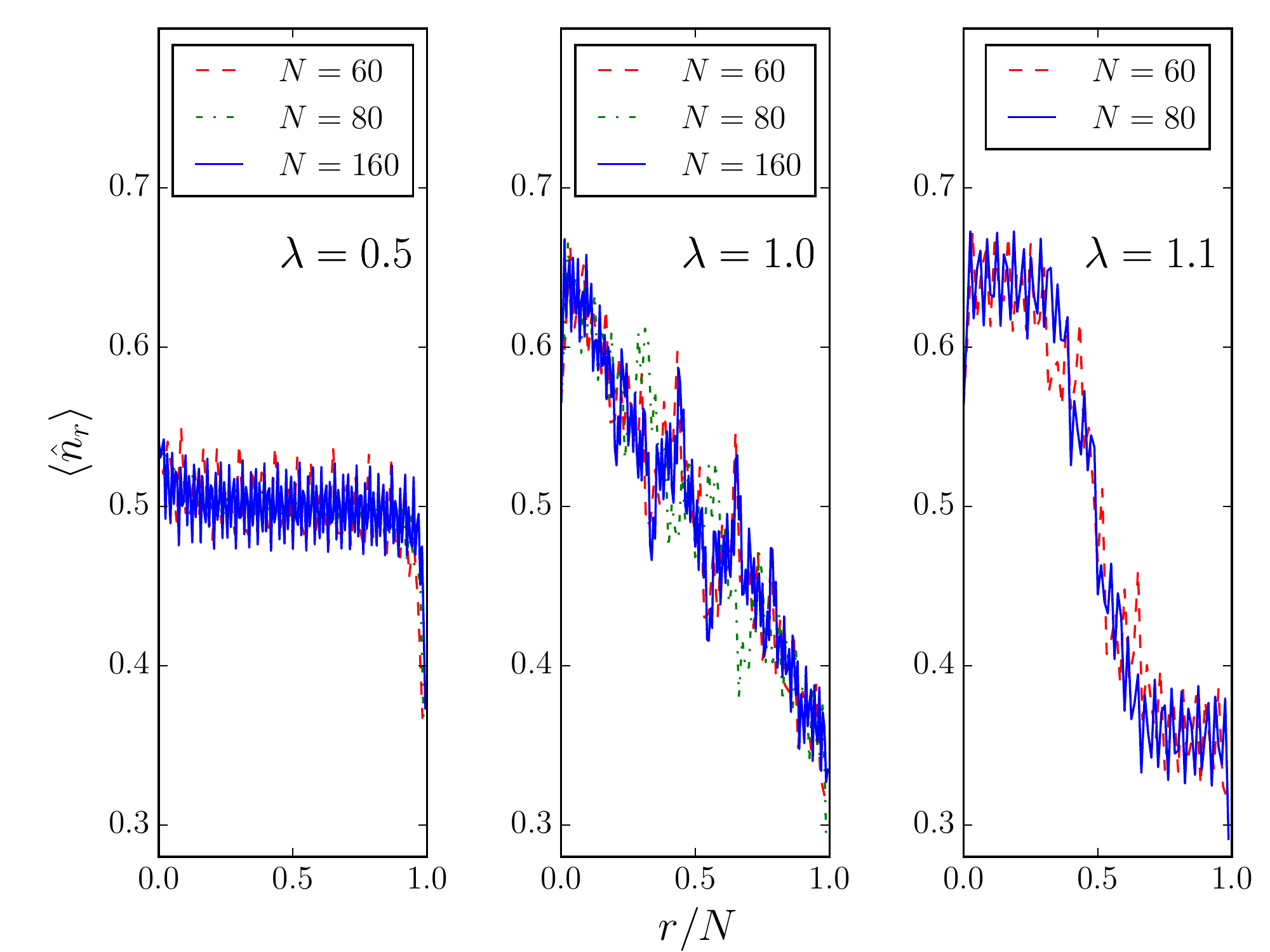} 
\caption{(color online) \textbf{Open system}: NESS particle density profile for the three regimes : delocalised ($\lambda=0.5<1$), critical ($\lambda=1$), localized ($\lambda=1.1>1$) for various system sizes. The particle density profile looks distinctly different in the three regimes. Parameters : $\beta=0.1$, $\mu_1=6, \mu_2=-6$, $\gamma=1$, $t_B=3$, $b=(\sqrt{5}-1)/2$ .    }
\label{NESS_n_plot}
\end{figure}

Note that, for $\lambda<1$, the single particle eigenfunctions are completely delocalized, hence $\Phi_{\alpha,\ell}^2 \sim N^{-1}$. Thus  $W\sim N^0$, thereby giving the ballistic scaling of current consistently. On the other hand, for $\lambda>1$, the single particle eigenfunctions are exponentially localized at some system site, so, $\Phi_{\alpha,1}^2\sim \Phi_{\alpha,N}^2\sim e^{-N}$. Thus, $W\sim e^{-N}$, thereby giving the exponential decay of current with system-size in this regime also consistently. Thus, the system size scaling of $W$ correctly gives the system-size scaling of currents at all regimes of the AAH model. This also shows that the current scaling with system size is independent of the details of the baths, and also independent of the type of particles (bosonic or fermionic).

Hence, the transport behavior of open AAH model is totally governed by the single particle eigenfunctions at the boundaries where the baths are attached. In the isolated thermodynamic limit, there are no boundaries, and the transport behavior is governed by the bulk properties. Looking at Eq.~\ref{sigma_simpf},~\ref{nxt},~\ref{Pxt}, and Eq.~\ref{W}, we see that there is no reason a priori that the isolated thermodynamic limit transport characterized by spread of a wavepacket, and the open system transport characterized by current scaling with system size, need to be consistent with each other in general. It nonetheless turns out that in the delocalized and the localized cases, they can indeed be shown to be consistent. The underlying reason for this is that, for these cases, the eigenstates contributing to transport have similar behavior in the bulk and at the boundaries. But, at the critical point, the eigenstates contributing to transport have different behavior at the bulk and at the boundaries. This leads to drastically different transport behavior in the isolated and in the open critical AAH model.

\subsubsection{NESS particle density profile}
Next, we look at the spatial particle density profile, $\langle \hat{n}_r \rangle$ vs $r$, in NESS in each of delocalized, critical and localized regimes when the two baths are at widely different chemical potentials. We find that the NESS spatial particle density profile (which is related to local chemical potential) behaves very differently in the three regimes (Fig.~\ref{NESS_n_plot}).  In the de-localized regime, we notice a flat profile, a hallmark of ballistic transport. In the critical regime, we see a continuous (almost linear) curve connecting the boundary densities. Such behaviour is typical of  diffusive systems. The localized regime shows a step-like profile and this has recently been reported for other models with localization  \cite{dhar2017,monthus}. Hence, this NESS physical quantity, which is potentially measurable with recent cutting-edge experiments \cite{expt0,expt1,expt2,expt3,expt5}, gives a clear real-space signature of localized, critical and delocalized phases. The energy profile (which is related to local temperature) has a similar behaviour.

\section{Conclusions}\label{conclusions} 
We have investigated the high temperature transport properties of the AAH model both in the isolated thermodynamic limit, and in the open system. We have found that the critical point of the AAH model has drastically different transport behavior in the two cases. In the isolated thermodynamic limit, spread of an initially localized wavepacket shows hints of super-diffusive behavior. The super-diffusive scaling exponent is non-universal and depends of the choice of the irrational number $b$. On the other hand, the open system NESS current $I$ scaling with system size $N$ is clearly sub-diffusive. There are two sub-diffusive exponents. One is $I\sim N^{-1.27\pm0.01}$, which is seen when system sizes are exactly the denominators of the rational approximants of $b$, while the other is   $I\sim N^{-1.4\pm0.05}$, which is the scaling for generic system sizes. We have shown that the current scaling with system size is entirely controlled by the system size scaling of  eigenfunctions at the boundaries where the baths are attached. Thus, the drastic difference between the isolated and the open system transport properties at the critical point is due to different behaviors of the eigenfunctions at the bulk and at the boundaries. 

We would like to point out that looking at the spread of correlations and measuring the current or conductance variation with system size are two different experiments done to characterize transport in many set-ups. Although not guaranteed, in many cases, the results of one experiment can be inferred from that of the other. We showed that at the critical point of the AAH model, this is not possible.

We have also looked that the NESS particle density profile of the open system connected to two baths at different chemical potentials. We have shown the the NESS particle density profile is distinctly different in the delocalized, critical and localized phases.

After submission of our work, a closely related work appeared \cite{vkv}, where very similar questions were explored using a phenomenological Lindblad quantum master equation approach. On the other hand, in our work, the baths are modelled by a microscopic quadratic Hamiltonians having infinite degrees of freedom, and results are calculated by fully exact NEGF method. This has no restrictions, for example, it is valid for arbitrary system-bath couplings and fully takes into account non-Markovianity. We find it remarkable that their work reproduces the same results (same scaling of current with system size) as ours. This, in our opinion, is important for the following reason. Because it matches with our results, it justifies the use of a phenomenological Lindblad quantum master equation approach which is often the most practical method for interacting systems \cite{Znidaric_sub_diff}.

\textit{Acknowledgements: } We would like to thank T. Kottos, H. Spohn, R. Moessner and A. Kundu for useful discussions. AD would like to thank support from the Indo-Israel joint research project No. 6- 8/2014(IC) and from the French Ministry of Education through the grant ANR (EDNHS). M. K. gratefully acknowledges the Ramanujan Fellowship SB/S2/RJN-114/2016 from the Science and Engineering Research Board (SERB), Department of Science and Technology, Government of India.

\bibliography{refAA}

%merlin.mbs apsrev4-1.bst 2010-07-25 4.21a (PWD, AO, DPC) hacked
%Control: key (0)
%Control: author (8) initials jnrlst
%Control: editor formatted (1) identically to author
%Control: production of article title (-1) disabled
%Control: page (0) single
%Control: year (1) truncated
%Control: production of eprint (0) enabled
\begin{thebibliography}{74}%
\makeatletter
\providecommand \@ifxundefined [1]{%
 \@ifx{#1\undefined}
}%
\providecommand \@ifnum [1]{%
 \ifnum #1\expandafter \@firstoftwo
 \else \expandafter \@secondoftwo
 \fi
}%
\providecommand \@ifx [1]{%
 \ifx #1\expandafter \@firstoftwo
 \else \expandafter \@secondoftwo
 \fi
}%
\providecommand \natexlab [1]{#1}%
\providecommand \enquote  [1]{``#1''}%
\providecommand \bibnamefont  [1]{#1}%
\providecommand \bibfnamefont [1]{#1}%
\providecommand \citenamefont [1]{#1}%
\providecommand \href@noop [0]{\@secondoftwo}%
\providecommand \href [0]{\begingroup \@sanitize@url \@href}%
\providecommand \@href[1]{\@@startlink{#1}\@@href}%
\providecommand \@@href[1]{\endgroup#1\@@endlink}%
\providecommand \@sanitize@url [0]{\catcode `\\12\catcode `\$12\catcode
  `\&12\catcode `\#12\catcode `\^12\catcode `\_12\catcode `\%12\relax}%
\providecommand \@@startlink[1]{}%
\providecommand \@@endlink[0]{}%
\providecommand \url  [0]{\begingroup\@sanitize@url \@url }%
\providecommand \@url [1]{\endgroup\@href {#1}{\urlprefix }}%
\providecommand \urlprefix  [0]{URL }%
\providecommand \Eprint [0]{\href }%
\providecommand \doibase [0]{http://dx.doi.org/}%
\providecommand \selectlanguage [0]{\@gobble}%
\providecommand \bibinfo  [0]{\@secondoftwo}%
\providecommand \bibfield  [0]{\@secondoftwo}%
\providecommand \translation [1]{[#1]}%
\providecommand \BibitemOpen [0]{}%
\providecommand \bibitemStop [0]{}%
\providecommand \bibitemNoStop [0]{.\EOS\space}%
\providecommand \EOS [0]{\spacefactor3000\relax}%
\providecommand \BibitemShut  [1]{\csname bibitem#1\endcsname}%
\let\auto@bib@innerbib\@empty
%</preamble>
\bibitem [{\citenamefont {Evers}\ and\ \citenamefont
  {Mirlin}(2008)}]{Mirlin_Evers}%
  \BibitemOpen
  \bibfield  {author} {\bibinfo {author} {\bibfnamefont {F.}~\bibnamefont
  {Evers}}\ and\ \bibinfo {author} {\bibfnamefont {A.~D.}\ \bibnamefont
  {Mirlin}},\ }\href {\doibase 10.1103/RevModPhys.80.1355} {\bibfield
  {journal} {\bibinfo  {journal} {Rev. Mod. Phys.}\ }\textbf {\bibinfo {volume}
  {80}},\ \bibinfo {pages} {1355} (\bibinfo {year} {2008})}\BibitemShut
  {NoStop}%
\bibitem [{\citenamefont {Lagendijk}\ \emph {et~al.}(2009)\citenamefont
  {Lagendijk}, \citenamefont {van Tiggelen},\ and\ \citenamefont
  {Wiersma}}]{Physics_Today_1}%
  \BibitemOpen
  \bibfield  {author} {\bibinfo {author} {\bibfnamefont {A.}~\bibnamefont
  {Lagendijk}}, \bibinfo {author} {\bibfnamefont {B.}~\bibnamefont {van
  Tiggelen}}, \ and\ \bibinfo {author} {\bibfnamefont {D.~S.}\ \bibnamefont
  {Wiersma}},\ }\href@noop {} {\bibfield  {journal} {\bibinfo  {journal}
  {Physics Today}\ }\textbf {\bibinfo {volume} {62}},\ \bibinfo {pages} {24}
  (\bibinfo {year} {2009})}\BibitemShut {NoStop}%
\bibitem [{\citenamefont {Aspect}\ and\ \citenamefont
  {Inguscio}(2009)}]{Physics_Today_2}%
  \BibitemOpen
  \bibfield  {author} {\bibinfo {author} {\bibfnamefont {A.}~\bibnamefont
  {Aspect}}\ and\ \bibinfo {author} {\bibfnamefont {M.}~\bibnamefont
  {Inguscio}},\ }\href@noop {} {\bibfield  {journal} {\bibinfo  {journal}
  {Physics Today}\ }\textbf {\bibinfo {volume} {62}},\ \bibinfo {pages} {30}
  (\bibinfo {year} {2009})}\BibitemShut {NoStop}%
\bibitem [{\citenamefont {Luitz}\ \emph {et~al.}(2016)\citenamefont {Luitz},
  \citenamefont {Laflorencie},\ and\ \citenamefont {Alet}}]{Griffiths4}%
  \BibitemOpen
  \bibfield  {author} {\bibinfo {author} {\bibfnamefont {D.~J.}\ \bibnamefont
  {Luitz}}, \bibinfo {author} {\bibfnamefont {N.}~\bibnamefont {Laflorencie}},
  \ and\ \bibinfo {author} {\bibfnamefont {F.}~\bibnamefont {Alet}},\ }\href
  {\doibase 10.1103/PhysRevB.93.060201} {\bibfield  {journal} {\bibinfo
  {journal} {Phys. Rev. B}\ }\textbf {\bibinfo {volume} {93}},\ \bibinfo
  {pages} {060201} (\bibinfo {year} {2016})}\BibitemShut {NoStop}%
\bibitem [{\citenamefont {Potter}\ \emph {et~al.}(2015)\citenamefont {Potter},
  \citenamefont {Vasseur},\ and\ \citenamefont {Parameswaran}}]{Griffiths3}%
  \BibitemOpen
  \bibfield  {author} {\bibinfo {author} {\bibfnamefont {A.~C.}\ \bibnamefont
  {Potter}}, \bibinfo {author} {\bibfnamefont {R.}~\bibnamefont {Vasseur}}, \
  and\ \bibinfo {author} {\bibfnamefont {S.~A.}\ \bibnamefont {Parameswaran}},\
  }\href {\doibase 10.1103/PhysRevX.5.031033} {\bibfield  {journal} {\bibinfo
  {journal} {Phys. Rev. X}\ }\textbf {\bibinfo {volume} {5}},\ \bibinfo {pages}
  {031033} (\bibinfo {year} {2015})}\BibitemShut {NoStop}%
\bibitem [{\citenamefont {Agarwal}\ \emph {et~al.}(2015)\citenamefont
  {Agarwal}, \citenamefont {Gopalakrishnan}, \citenamefont {Knap},
  \citenamefont {M\"uller},\ and\ \citenamefont {Demler}}]{Griffiths2}%
  \BibitemOpen
  \bibfield  {author} {\bibinfo {author} {\bibfnamefont {K.}~\bibnamefont
  {Agarwal}}, \bibinfo {author} {\bibfnamefont {S.}~\bibnamefont
  {Gopalakrishnan}}, \bibinfo {author} {\bibfnamefont {M.}~\bibnamefont
  {Knap}}, \bibinfo {author} {\bibfnamefont {M.}~\bibnamefont {M\"uller}}, \
  and\ \bibinfo {author} {\bibfnamefont {E.}~\bibnamefont {Demler}},\ }\href
  {\doibase 10.1103/PhysRevLett.114.160401} {\bibfield  {journal} {\bibinfo
  {journal} {Phys. Rev. Lett.}\ }\textbf {\bibinfo {volume} {114}},\ \bibinfo
  {pages} {160401} (\bibinfo {year} {2015})}\BibitemShut {NoStop}%
\bibitem [{\citenamefont {Vosk}\ \emph {et~al.}(2015)\citenamefont {Vosk},
  \citenamefont {Huse},\ and\ \citenamefont {Altman}}]{Griffiths1}%
  \BibitemOpen
  \bibfield  {author} {\bibinfo {author} {\bibfnamefont {R.}~\bibnamefont
  {Vosk}}, \bibinfo {author} {\bibfnamefont {D.~A.}\ \bibnamefont {Huse}}, \
  and\ \bibinfo {author} {\bibfnamefont {E.}~\bibnamefont {Altman}},\ }\href
  {\doibase 10.1103/PhysRevX.5.031032} {\bibfield  {journal} {\bibinfo
  {journal} {Phys. Rev. X}\ }\textbf {\bibinfo {volume} {5}},\ \bibinfo {pages}
  {031032} (\bibinfo {year} {2015})}\BibitemShut {NoStop}%
\bibitem [{\citenamefont {Aubry}\ and\ \citenamefont {Andre}(1980)}]{aa1}%
  \BibitemOpen
  \bibfield  {author} {\bibinfo {author} {\bibfnamefont {S.}~\bibnamefont
  {Aubry}}\ and\ \bibinfo {author} {\bibfnamefont {G.}~\bibnamefont {Andre}},\
  }\href@noop {} {\bibfield  {journal} {\bibinfo  {journal} {Ann. Israel Phys.
  Soc}\ }\textbf {\bibinfo {volume} {3}},\ \bibinfo {pages} {18} (\bibinfo
  {year} {1980})}\BibitemShut {NoStop}%
\bibitem [{\citenamefont {Harper}(1955)}]{harper}%
  \BibitemOpen
  \bibfield  {author} {\bibinfo {author} {\bibfnamefont {P.~G.}\ \bibnamefont
  {Harper}},\ }\href@noop {} {\bibfield  {journal} {\bibinfo  {journal} {Proc.
  Roy. Soc. London, Ser. A}\ ,\ \bibinfo {pages} {874}} (\bibinfo {year}
  {1955})}\BibitemShut {NoStop}%
\bibitem [{\citenamefont {Hofstadter}(1976)}]{fractal1}%
  \BibitemOpen
  \bibfield  {author} {\bibinfo {author} {\bibfnamefont {D.~R.}\ \bibnamefont
  {Hofstadter}},\ }\href@noop {} {\bibfield  {journal} {\bibinfo  {journal}
  {Phys. Rev. B}\ }\textbf {\bibinfo {volume} {14}},\ \bibinfo {pages} {2239}
  (\bibinfo {year} {1976})}\BibitemShut {NoStop}%
\bibitem [{\citenamefont {Ketzmerick}\ \emph {et~al.}(1998)\citenamefont
  {Ketzmerick}, \citenamefont {Kruse}, \citenamefont {Steinbach},\ and\
  \citenamefont {Geisel}}]{fractal2}%
  \BibitemOpen
  \bibfield  {author} {\bibinfo {author} {\bibfnamefont {R.}~\bibnamefont
  {Ketzmerick}}, \bibinfo {author} {\bibfnamefont {K.}~\bibnamefont {Kruse}},
  \bibinfo {author} {\bibfnamefont {F.}~\bibnamefont {Steinbach}}, \ and\
  \bibinfo {author} {\bibfnamefont {T.}~\bibnamefont {Geisel}},\ }\href@noop {}
  {\bibfield  {journal} {\bibinfo  {journal} {Phys. Rev. B}\ }\textbf {\bibinfo
  {volume} {58}},\ \bibinfo {pages} {9881} (\bibinfo {year}
  {1998})}\BibitemShut {NoStop}%
\bibitem [{\citenamefont {Ostlund}\ \emph {et~al.}(1983)\citenamefont
  {Ostlund}, \citenamefont {Pandit}, \citenamefont {Rand}, \citenamefont
  {Schellnhuber},\ and\ \citenamefont {Siggia}}]{pandit83}%
  \BibitemOpen
  \bibfield  {author} {\bibinfo {author} {\bibfnamefont {S.}~\bibnamefont
  {Ostlund}}, \bibinfo {author} {\bibfnamefont {R.}~\bibnamefont {Pandit}},
  \bibinfo {author} {\bibfnamefont {D.}~\bibnamefont {Rand}}, \bibinfo {author}
  {\bibfnamefont {H.~J.}\ \bibnamefont {Schellnhuber}}, \ and\ \bibinfo
  {author} {\bibfnamefont {E.~D.}\ \bibnamefont {Siggia}},\ }\href@noop {}
  {\bibfield  {journal} {\bibinfo  {journal} {Phys. Rev. Lett.}\ }\textbf
  {\bibinfo {volume} {50}},\ \bibinfo {pages} {1873} (\bibinfo {year}
  {1983})}\BibitemShut {NoStop}%
\bibitem [{\citenamefont {Avila}\ and\ \citenamefont
  {Jitomirskaya}(2009)}]{math1}%
  \BibitemOpen
  \bibfield  {author} {\bibinfo {author} {\bibfnamefont {A.}~\bibnamefont
  {Avila}}\ and\ \bibinfo {author} {\bibfnamefont {S.}~\bibnamefont
  {Jitomirskaya}},\ }\href@noop {} {\bibfield  {journal} {\bibinfo  {journal}
  {Annals of Mathematics}\ }\textbf {\bibinfo {volume} {170}},\ \bibinfo
  {pages} {303} (\bibinfo {year} {2009})}\BibitemShut {NoStop}%
\bibitem [{\citenamefont {Last}(1994)}]{math2}%
  \BibitemOpen
  \bibfield  {author} {\bibinfo {author} {\bibfnamefont {Y.}~\bibnamefont
  {Last}},\ }\href@noop {} {\bibfield  {journal} {\bibinfo  {journal}
  {Communications in Mathematical Physics}\ }\textbf {\bibinfo {volume}
  {164}},\ \bibinfo {pages} {421} (\bibinfo {year} {1994})}\BibitemShut
  {NoStop}%
\bibitem [{\citenamefont {Yang}\ \emph {et~al.}(2017)\citenamefont {Yang},
  \citenamefont {Wang}, \citenamefont {Wang}, \citenamefont {Xianlong},\ and\
  \citenamefont {Chen}}]{Yang2017}%
  \BibitemOpen
  \bibfield  {author} {\bibinfo {author} {\bibfnamefont {C.}~\bibnamefont
  {Yang}}, \bibinfo {author} {\bibfnamefont {Y.}~\bibnamefont {Wang}}, \bibinfo
  {author} {\bibfnamefont {P.}~\bibnamefont {Wang}}, \bibinfo {author}
  {\bibfnamefont {G.}~\bibnamefont {Xianlong}}, \ and\ \bibinfo {author}
  {\bibfnamefont {S.}~\bibnamefont {Chen}},\ }\href@noop {} {\bibfield
  {journal} {\bibinfo  {journal} {arXiv:1703.07489}\ } (\bibinfo {year}
  {2017})}\BibitemShut {NoStop}%
\bibitem [{\citenamefont {Lev}\ \emph {et~al.}(2017)\citenamefont {Lev},
  \citenamefont {Kennes}, \citenamefont {Klöckner}, \citenamefont {Reichman},\
  and\ \citenamefont {Karrasch}}]{david2017}%
  \BibitemOpen
  \bibfield  {author} {\bibinfo {author} {\bibfnamefont {Y.~B.}\ \bibnamefont
  {Lev}}, \bibinfo {author} {\bibfnamefont {D.~M.}\ \bibnamefont {Kennes}},
  \bibinfo {author} {\bibfnamefont {C.}~\bibnamefont {Klöckner}}, \bibinfo
  {author} {\bibfnamefont {D.~R.}\ \bibnamefont {Reichman}}, \ and\ \bibinfo
  {author} {\bibfnamefont {C.}~\bibnamefont {Karrasch}},\ }\href@noop {}
  {\bibfield  {journal} {\bibinfo  {journal} {arXiv:1702.04349}\ } (\bibinfo
  {year} {2017})}\BibitemShut {NoStop}%
\bibitem [{\citenamefont {Naldesi}\ \emph {et~al.}(2016)\citenamefont
  {Naldesi}, \citenamefont {Ercolessi},\ and\ \citenamefont
  {Roscilde}}]{Naldesi2016}%
  \BibitemOpen
  \bibfield  {author} {\bibinfo {author} {\bibfnamefont {P.}~\bibnamefont
  {Naldesi}}, \bibinfo {author} {\bibfnamefont {E.}~\bibnamefont {Ercolessi}},
  \ and\ \bibinfo {author} {\bibfnamefont {T.}~\bibnamefont {Roscilde}},\
  }\href@noop {} {\bibfield  {journal} {\bibinfo  {journal} {SciPost Phys.}\
  }\textbf {\bibinfo {volume} {1}},\ \bibinfo {pages} {010} (\bibinfo {year}
  {2016})}\BibitemShut {NoStop}%
\bibitem [{\citenamefont {Wang}\ \emph {et~al.}(2017)\citenamefont {Wang},
  \citenamefont {Liu}, \citenamefont {Chen},\ and\ \citenamefont
  {Zhang}}]{AAH2017}%
  \BibitemOpen
  \bibfield  {author} {\bibinfo {author} {\bibfnamefont {L.}~\bibnamefont
  {Wang}}, \bibinfo {author} {\bibfnamefont {N.}~\bibnamefont {Liu}}, \bibinfo
  {author} {\bibfnamefont {S.}~\bibnamefont {Chen}}, \ and\ \bibinfo {author}
  {\bibfnamefont {Y.}~\bibnamefont {Zhang}},\ }\href {\doibase
  10.1103/PhysRevA.95.013619} {\bibfield  {journal} {\bibinfo  {journal} {Phys.
  Rev. A}\ }\textbf {\bibinfo {volume} {95}},\ \bibinfo {pages} {013619}
  (\bibinfo {year} {2017})}\BibitemShut {NoStop}%
\bibitem [{\citenamefont {Deng}\ \emph {et~al.}(2016)\citenamefont {Deng},
  \citenamefont {Ganeshan}, \citenamefont {Li}, \citenamefont {Modak},
  \citenamefont {Mukerjee},\ and\ \citenamefont {Pixley}}]{AAH1}%
  \BibitemOpen
  \bibfield  {author} {\bibinfo {author} {\bibfnamefont {D.-L.}\ \bibnamefont
  {Deng}}, \bibinfo {author} {\bibfnamefont {S.}~\bibnamefont {Ganeshan}},
  \bibinfo {author} {\bibfnamefont {X.}~\bibnamefont {Li}}, \bibinfo {author}
  {\bibfnamefont {R.}~\bibnamefont {Modak}}, \bibinfo {author} {\bibfnamefont
  {S.}~\bibnamefont {Mukerjee}}, \ and\ \bibinfo {author} {\bibfnamefont
  {J.~H.}\ \bibnamefont {Pixley}},\ }\href@noop {} {\bibfield  {journal}
  {\bibinfo  {journal} {arXiv:1612.00976}\ } (\bibinfo {year}
  {2016})}\BibitemShut {NoStop}%
\bibitem [{\citenamefont {Li}\ \emph {et~al.}(2016{\natexlab{a}})\citenamefont
  {Li}, \citenamefont {Pixley}, \citenamefont {Deng}, \citenamefont
  {Ganeshan},\ and\ \citenamefont {Das~Sarma}}]{AAH2}%
  \BibitemOpen
  \bibfield  {author} {\bibinfo {author} {\bibfnamefont {X.}~\bibnamefont
  {Li}}, \bibinfo {author} {\bibfnamefont {J.~H.}\ \bibnamefont {Pixley}},
  \bibinfo {author} {\bibfnamefont {D.-L.}\ \bibnamefont {Deng}}, \bibinfo
  {author} {\bibfnamefont {S.}~\bibnamefont {Ganeshan}}, \ and\ \bibinfo
  {author} {\bibfnamefont {S.}~\bibnamefont {Das~Sarma}},\ }\href@noop {}
  {\bibfield  {journal} {\bibinfo  {journal} {Phys. Rev. B}\ }\textbf {\bibinfo
  {volume} {93}},\ \bibinfo {pages} {184204} (\bibinfo {year}
  {2016}{\natexlab{a}})}\BibitemShut {NoStop}%
\bibitem [{\citenamefont {Zeng}\ \emph {et~al.}(2016)\citenamefont {Zeng},
  \citenamefont {Chen},\ and\ \citenamefont {L\"u}}]{AAH3}%
  \BibitemOpen
  \bibfield  {author} {\bibinfo {author} {\bibfnamefont {Q.-B.}\ \bibnamefont
  {Zeng}}, \bibinfo {author} {\bibfnamefont {S.}~\bibnamefont {Chen}}, \ and\
  \bibinfo {author} {\bibfnamefont {R.}~\bibnamefont {L\"u}},\ }\href@noop {}
  {\bibfield  {journal} {\bibinfo  {journal} {Phys. Rev. B}\ }\textbf {\bibinfo
  {volume} {94}},\ \bibinfo {pages} {125408} (\bibinfo {year}
  {2016})}\BibitemShut {NoStop}%
\bibitem [{\citenamefont {Modak}\ \emph {et~al.}(2016)\citenamefont {Modak},
  \citenamefont {Mukerjee}, \citenamefont {Yuzbashyan},\ and\ \citenamefont
  {Shastry}}]{AAH4}%
  \BibitemOpen
  \bibfield  {author} {\bibinfo {author} {\bibfnamefont {R.}~\bibnamefont
  {Modak}}, \bibinfo {author} {\bibfnamefont {S.}~\bibnamefont {Mukerjee}},
  \bibinfo {author} {\bibfnamefont {E.~A.}\ \bibnamefont {Yuzbashyan}}, \ and\
  \bibinfo {author} {\bibfnamefont {B.~S.}\ \bibnamefont {Shastry}},\
  }\href@noop {} {\bibfield  {journal} {\bibinfo  {journal} {New Journal of
  Physics}\ }\textbf {\bibinfo {volume} {18}},\ \bibinfo {pages} {033010}
  (\bibinfo {year} {2016})}\BibitemShut {NoStop}%
\bibitem [{\citenamefont {Li}\ \emph {et~al.}(2016{\natexlab{b}})\citenamefont
  {Li}, \citenamefont {Pixley}, \citenamefont {Deng}, \citenamefont
  {Ganeshan},\ and\ \citenamefont {Das~Sarma}}]{AAH5}%
  \BibitemOpen
  \bibfield  {author} {\bibinfo {author} {\bibfnamefont {X.}~\bibnamefont
  {Li}}, \bibinfo {author} {\bibfnamefont {J.~H.}\ \bibnamefont {Pixley}},
  \bibinfo {author} {\bibfnamefont {D.-L.}\ \bibnamefont {Deng}}, \bibinfo
  {author} {\bibfnamefont {S.}~\bibnamefont {Ganeshan}}, \ and\ \bibinfo
  {author} {\bibfnamefont {S.}~\bibnamefont {Das~Sarma}},\ }\href {\doibase
  10.1103/PhysRevB.93.184204} {\bibfield  {journal} {\bibinfo  {journal} {Phys.
  Rev. B}\ }\textbf {\bibinfo {volume} {93}},\ \bibinfo {pages} {184204}
  (\bibinfo {year} {2016}{\natexlab{b}})}\BibitemShut {NoStop}%
\bibitem [{\citenamefont {Ray}\ \emph {et~al.}(2016)\citenamefont {Ray},
  \citenamefont {Pandey}, \citenamefont {Ghosh},\ and\ \citenamefont
  {Sinha}}]{AAH6}%
  \BibitemOpen
  \bibfield  {author} {\bibinfo {author} {\bibfnamefont {S.}~\bibnamefont
  {Ray}}, \bibinfo {author} {\bibfnamefont {M.}~\bibnamefont {Pandey}},
  \bibinfo {author} {\bibfnamefont {A.}~\bibnamefont {Ghosh}}, \ and\ \bibinfo
  {author} {\bibfnamefont {S.}~\bibnamefont {Sinha}},\ }\href@noop {}
  {\bibfield  {journal} {\bibinfo  {journal} {New Journal of Physics}\ }\textbf
  {\bibinfo {volume} {18}},\ \bibinfo {pages} {013013} (\bibinfo {year}
  {2016})}\BibitemShut {NoStop}%
\bibitem [{\citenamefont {Saha}\ \emph {et~al.}(2016)\citenamefont {Saha},
  \citenamefont {Maiti},\ and\ \citenamefont {Karmakar}}]{AAH7}%
  \BibitemOpen
  \bibfield  {author} {\bibinfo {author} {\bibfnamefont {S.}~\bibnamefont
  {Saha}}, \bibinfo {author} {\bibfnamefont {S.~K.}\ \bibnamefont {Maiti}}, \
  and\ \bibinfo {author} {\bibfnamefont {S.}~\bibnamefont {Karmakar}},\
  }\href@noop {} {\bibfield  {journal} {\bibinfo  {journal} {Physica E}\
  }\textbf {\bibinfo {volume} {83}},\ \bibinfo {pages} {358 } (\bibinfo {year}
  {2016})}\BibitemShut {NoStop}%
\bibitem [{\citenamefont {Mastropietro}(2015)}]{AAH8}%
  \BibitemOpen
  \bibfield  {author} {\bibinfo {author} {\bibfnamefont {V.}~\bibnamefont
  {Mastropietro}},\ }\href@noop {} {\bibfield  {journal} {\bibinfo  {journal}
  {Phys. Rev. Lett.}\ }\textbf {\bibinfo {volume} {115}},\ \bibinfo {pages}
  {180401} (\bibinfo {year} {2015})}\BibitemShut {NoStop}%
\bibitem [{\citenamefont {Li}\ \emph {et~al.}(2015)\citenamefont {Li},
  \citenamefont {Ganeshan}, \citenamefont {Pixley},\ and\ \citenamefont
  {Das~Sarma}}]{AAH9}%
  \BibitemOpen
  \bibfield  {author} {\bibinfo {author} {\bibfnamefont {X.}~\bibnamefont
  {Li}}, \bibinfo {author} {\bibfnamefont {S.}~\bibnamefont {Ganeshan}},
  \bibinfo {author} {\bibfnamefont {J.~H.}\ \bibnamefont {Pixley}}, \ and\
  \bibinfo {author} {\bibfnamefont {S.}~\bibnamefont {Das~Sarma}},\ }\href
  {\doibase 10.1103/PhysRevLett.115.186601} {\bibfield  {journal} {\bibinfo
  {journal} {Phys. Rev. Lett.}\ }\textbf {\bibinfo {volume} {115}},\ \bibinfo
  {pages} {186601} (\bibinfo {year} {2015})}\BibitemShut {NoStop}%
\bibitem [{\citenamefont {Modak}\ and\ \citenamefont {Mukerjee}(2015)}]{AAH10}%
  \BibitemOpen
  \bibfield  {author} {\bibinfo {author} {\bibfnamefont {R.}~\bibnamefont
  {Modak}}\ and\ \bibinfo {author} {\bibfnamefont {S.}~\bibnamefont
  {Mukerjee}},\ }\href@noop {} {\bibfield  {journal} {\bibinfo  {journal}
  {Phys. Rev. Lett.}\ }\textbf {\bibinfo {volume} {115}},\ \bibinfo {pages}
  {230401} (\bibinfo {year} {2015})}\BibitemShut {NoStop}%
\bibitem [{\citenamefont {Ganeshan}\ \emph {et~al.}(2015)\citenamefont
  {Ganeshan}, \citenamefont {Pixley},\ and\ \citenamefont {Das~Sarma}}]{AAH11}%
  \BibitemOpen
  \bibfield  {author} {\bibinfo {author} {\bibfnamefont {S.}~\bibnamefont
  {Ganeshan}}, \bibinfo {author} {\bibfnamefont {J.~H.}\ \bibnamefont
  {Pixley}}, \ and\ \bibinfo {author} {\bibfnamefont {S.}~\bibnamefont
  {Das~Sarma}},\ }\href@noop {} {\bibfield  {journal} {\bibinfo  {journal}
  {Phys. Rev. Lett.}\ }\textbf {\bibinfo {volume} {114}},\ \bibinfo {pages}
  {146601} (\bibinfo {year} {2015})}\BibitemShut {NoStop}%
\bibitem [{\citenamefont {Morales-Molina}\ \emph {et~al.}(2014)\citenamefont
  {Morales-Molina}, \citenamefont {Doerner}, \citenamefont {Danieli},\ and\
  \citenamefont {Flach}}]{AAH12}%
  \BibitemOpen
  \bibfield  {author} {\bibinfo {author} {\bibfnamefont {L.}~\bibnamefont
  {Morales-Molina}}, \bibinfo {author} {\bibfnamefont {E.}~\bibnamefont
  {Doerner}}, \bibinfo {author} {\bibfnamefont {C.}~\bibnamefont {Danieli}}, \
  and\ \bibinfo {author} {\bibfnamefont {S.}~\bibnamefont {Flach}},\
  }\href@noop {} {\bibfield  {journal} {\bibinfo  {journal} {Phys. Rev. A}\
  }\textbf {\bibinfo {volume} {90}},\ \bibinfo {pages} {043630} (\bibinfo
  {year} {2014})}\BibitemShut {NoStop}%
\bibitem [{\citenamefont {Shen}\ \emph {et~al.}(2014)\citenamefont {Shen},
  \citenamefont {Yi},\ and\ \citenamefont {Oh}}]{AAH13}%
  \BibitemOpen
  \bibfield  {author} {\bibinfo {author} {\bibfnamefont {H.~Z.}\ \bibnamefont
  {Shen}}, \bibinfo {author} {\bibfnamefont {X.~X.}\ \bibnamefont {Yi}}, \ and\
  \bibinfo {author} {\bibfnamefont {C.~H.}\ \bibnamefont {Oh}},\ }\href@noop {}
  {\bibfield  {journal} {\bibinfo  {journal} {Journal of Physics B: Atomic,
  Molecular and Optical Physics}\ }\textbf {\bibinfo {volume} {47}},\ \bibinfo
  {pages} {085501} (\bibinfo {year} {2014})}\BibitemShut {NoStop}%
\bibitem [{\citenamefont {Ro\'osz}\ \emph {et~al.}(2014)\citenamefont
  {Ro\'osz}, \citenamefont {Divakaran}, \citenamefont {Rieger},\ and\
  \citenamefont {Igl\'oi}}]{AAH14}%
  \BibitemOpen
  \bibfield  {author} {\bibinfo {author} {\bibfnamefont {G.}~\bibnamefont
  {Ro\'osz}}, \bibinfo {author} {\bibfnamefont {U.}~\bibnamefont {Divakaran}},
  \bibinfo {author} {\bibfnamefont {H.}~\bibnamefont {Rieger}}, \ and\ \bibinfo
  {author} {\bibfnamefont {F.}~\bibnamefont {Igl\'oi}},\ }\href@noop {}
  {\bibfield  {journal} {\bibinfo  {journal} {Phys. Rev. B}\ }\textbf {\bibinfo
  {volume} {90}},\ \bibinfo {pages} {184202} (\bibinfo {year}
  {2014})}\BibitemShut {NoStop}%
\bibitem [{\citenamefont {Guo}\ \emph {et~al.}(2014)\citenamefont {Guo},
  \citenamefont {Xie},\ and\ \citenamefont {Sun}}]{AAH15}%
  \BibitemOpen
  \bibfield  {author} {\bibinfo {author} {\bibfnamefont {A.-M.}\ \bibnamefont
  {Guo}}, \bibinfo {author} {\bibfnamefont {X.~C.}\ \bibnamefont {Xie}}, \ and\
  \bibinfo {author} {\bibfnamefont {Q.-f.}\ \bibnamefont {Sun}},\ }\href@noop
  {} {\bibfield  {journal} {\bibinfo  {journal} {Phys. Rev. B}\ }\textbf
  {\bibinfo {volume} {89}},\ \bibinfo {pages} {075434} (\bibinfo {year}
  {2014})}\BibitemShut {NoStop}%
\bibitem [{\citenamefont {Radosavljevi\'c}\ \emph {et~al.}(2014)\citenamefont
  {Radosavljevi\'c}, \citenamefont {Gligori\'c}, \citenamefont {Maluckov},\
  and\ \citenamefont {Stepi\'c}}]{AAH16}%
  \BibitemOpen
  \bibfield  {author} {\bibinfo {author} {\bibfnamefont {A.}~\bibnamefont
  {Radosavljevi\'c}}, \bibinfo {author} {\bibfnamefont {G.}~\bibnamefont
  {Gligori\'c}}, \bibinfo {author} {\bibfnamefont {A.}~\bibnamefont
  {Maluckov}}, \ and\ \bibinfo {author} {\bibfnamefont {M.}~\bibnamefont
  {Stepi\'c}},\ }\href@noop {} {\bibfield  {journal} {\bibinfo  {journal}
  {Journal of Optics}\ }\textbf {\bibinfo {volume} {16}},\ \bibinfo {pages}
  {025201} (\bibinfo {year} {2014})}\BibitemShut {NoStop}%
\bibitem [{\citenamefont {Iyer}\ \emph {et~al.}(2013)\citenamefont {Iyer},
  \citenamefont {Oganesyan}, \citenamefont {Refael},\ and\ \citenamefont
  {Huse}}]{AAH17}%
  \BibitemOpen
  \bibfield  {author} {\bibinfo {author} {\bibfnamefont {S.}~\bibnamefont
  {Iyer}}, \bibinfo {author} {\bibfnamefont {V.}~\bibnamefont {Oganesyan}},
  \bibinfo {author} {\bibfnamefont {G.}~\bibnamefont {Refael}}, \ and\ \bibinfo
  {author} {\bibfnamefont {D.~A.}\ \bibnamefont {Huse}},\ }\href@noop {}
  {\bibfield  {journal} {\bibinfo  {journal} {Phys. Rev. B}\ }\textbf {\bibinfo
  {volume} {87}},\ \bibinfo {pages} {134202} (\bibinfo {year}
  {2013})}\BibitemShut {NoStop}%
\bibitem [{\citenamefont {Larcher}\ \emph {et~al.}(2012)\citenamefont
  {Larcher}, \citenamefont {Laptyeva}, \citenamefont {Bodyfelt}, \citenamefont
  {Dalfovo}, \citenamefont {Modugno},\ and\ \citenamefont {Flach}}]{AAH18}%
  \BibitemOpen
  \bibfield  {author} {\bibinfo {author} {\bibfnamefont {M.}~\bibnamefont
  {Larcher}}, \bibinfo {author} {\bibfnamefont {T.~V.}\ \bibnamefont
  {Laptyeva}}, \bibinfo {author} {\bibfnamefont {J.~D.}\ \bibnamefont
  {Bodyfelt}}, \bibinfo {author} {\bibfnamefont {F.}~\bibnamefont {Dalfovo}},
  \bibinfo {author} {\bibfnamefont {M.}~\bibnamefont {Modugno}}, \ and\
  \bibinfo {author} {\bibfnamefont {S.}~\bibnamefont {Flach}},\ }\href@noop {}
  {\bibfield  {journal} {\bibinfo  {journal} {New Journal of Physics}\ }\textbf
  {\bibinfo {volume} {14}},\ \bibinfo {pages} {103036} (\bibinfo {year}
  {2012})}\BibitemShut {NoStop}%
\bibitem [{\citenamefont {Tezuka}\ and\ \citenamefont
  {Garc\'{\i}a-Garc\'{\i}a}(2012)}]{AAH19}%
  \BibitemOpen
  \bibfield  {author} {\bibinfo {author} {\bibfnamefont {M.}~\bibnamefont
  {Tezuka}}\ and\ \bibinfo {author} {\bibfnamefont {A.~M.}\ \bibnamefont
  {Garc\'{\i}a-Garc\'{\i}a}},\ }\href {\doibase 10.1103/PhysRevA.85.031602}
  {\bibfield  {journal} {\bibinfo  {journal} {Phys. Rev. A}\ }\textbf {\bibinfo
  {volume} {85}},\ \bibinfo {pages} {031602} (\bibinfo {year}
  {2012})}\BibitemShut {NoStop}%
\bibitem [{\citenamefont {L{\"u}schen}\ \emph {et~al.}(2016)\citenamefont
  {L{\"u}schen}, \citenamefont {Bordia}, \citenamefont {Scherg}, \citenamefont
  {Alet}, \citenamefont {Altman}, \citenamefont {Schneider},\ and\
  \citenamefont {Bloch}}]{expt0}%
  \BibitemOpen
  \bibfield  {author} {\bibinfo {author} {\bibfnamefont {H.~P.}\ \bibnamefont
  {L{\"u}schen}}, \bibinfo {author} {\bibfnamefont {P.}~\bibnamefont {Bordia}},
  \bibinfo {author} {\bibfnamefont {S.}~\bibnamefont {Scherg}}, \bibinfo
  {author} {\bibfnamefont {F.}~\bibnamefont {Alet}}, \bibinfo {author}
  {\bibfnamefont {E.}~\bibnamefont {Altman}}, \bibinfo {author} {\bibfnamefont
  {U.}~\bibnamefont {Schneider}}, \ and\ \bibinfo {author} {\bibfnamefont
  {I.}~\bibnamefont {Bloch}},\ }\href@noop {} {\bibfield  {journal} {\bibinfo
  {journal} {arXiv:1612.07173}\ } (\bibinfo {year} {2016})}\BibitemShut
  {NoStop}%
\bibitem [{\citenamefont {L\"uschen}\ \emph {et~al.}(2017)\citenamefont
  {L\"uschen}, \citenamefont {Bordia}, \citenamefont {Hodgman}, \citenamefont
  {Schreiber}, \citenamefont {Sarkar}, \citenamefont {Daley}, \citenamefont
  {Fischer}, \citenamefont {Altman}, \citenamefont {Bloch},\ and\ \citenamefont
  {Schneider}}]{expt1}%
  \BibitemOpen
  \bibfield  {author} {\bibinfo {author} {\bibfnamefont {H.~P.}\ \bibnamefont
  {L\"uschen}}, \bibinfo {author} {\bibfnamefont {P.}~\bibnamefont {Bordia}},
  \bibinfo {author} {\bibfnamefont {S.~S.}\ \bibnamefont {Hodgman}}, \bibinfo
  {author} {\bibfnamefont {M.}~\bibnamefont {Schreiber}}, \bibinfo {author}
  {\bibfnamefont {S.}~\bibnamefont {Sarkar}}, \bibinfo {author} {\bibfnamefont
  {A.~J.}\ \bibnamefont {Daley}}, \bibinfo {author} {\bibfnamefont {M.~H.}\
  \bibnamefont {Fischer}}, \bibinfo {author} {\bibfnamefont {E.}~\bibnamefont
  {Altman}}, \bibinfo {author} {\bibfnamefont {I.}~\bibnamefont {Bloch}}, \
  and\ \bibinfo {author} {\bibfnamefont {U.}~\bibnamefont {Schneider}},\ }\href
  {\doibase 10.1103/PhysRevX.7.011034} {\bibfield  {journal} {\bibinfo
  {journal} {Phys. Rev. X}\ }\textbf {\bibinfo {volume} {7}},\ \bibinfo {pages}
  {011034} (\bibinfo {year} {2017})}\BibitemShut {NoStop}%
\bibitem [{\citenamefont {Schreiber}\ \emph {et~al.}(2015)\citenamefont
  {Schreiber}, \citenamefont {Hodgman}, \citenamefont {Bordia}, \citenamefont
  {L{\"u}schen}, \citenamefont {Fischer}, \citenamefont {Vosk}, \citenamefont
  {Altman}, \citenamefont {Schneider},\ and\ \citenamefont {Bloch}}]{expt2}%
  \BibitemOpen
  \bibfield  {author} {\bibinfo {author} {\bibfnamefont {M.}~\bibnamefont
  {Schreiber}}, \bibinfo {author} {\bibfnamefont {S.~S.}\ \bibnamefont
  {Hodgman}}, \bibinfo {author} {\bibfnamefont {P.}~\bibnamefont {Bordia}},
  \bibinfo {author} {\bibfnamefont {H.~P.}\ \bibnamefont {L{\"u}schen}},
  \bibinfo {author} {\bibfnamefont {M.~H.}\ \bibnamefont {Fischer}}, \bibinfo
  {author} {\bibfnamefont {R.}~\bibnamefont {Vosk}}, \bibinfo {author}
  {\bibfnamefont {E.}~\bibnamefont {Altman}}, \bibinfo {author} {\bibfnamefont
  {U.}~\bibnamefont {Schneider}}, \ and\ \bibinfo {author} {\bibfnamefont
  {I.}~\bibnamefont {Bloch}},\ }\href {\doibase 10.1126/science.aaa7432}
  {\bibfield  {journal} {\bibinfo  {journal} {Science}\ }\textbf {\bibinfo
  {volume} {349}},\ \bibinfo {pages} {842} (\bibinfo {year}
  {2015})}\BibitemShut {NoStop}%
\bibitem [{\citenamefont {\'Errico}\ \emph {et~al.}(2013)\citenamefont
  {\'Errico}, \citenamefont {Moratti}, \citenamefont {Lucioni}, \citenamefont
  {Tanzi}, \citenamefont {Deissler}, \citenamefont {Inguscio}, \citenamefont
  {Modugno}, \citenamefont {Plenio},\ and\ \citenamefont {Caruso}}]{expt3}%
  \BibitemOpen
  \bibfield  {author} {\bibinfo {author} {\bibfnamefont {C.~D.}\ \bibnamefont
  {\'Errico}}, \bibinfo {author} {\bibfnamefont {M.}~\bibnamefont {Moratti}},
  \bibinfo {author} {\bibfnamefont {E.}~\bibnamefont {Lucioni}}, \bibinfo
  {author} {\bibfnamefont {L.}~\bibnamefont {Tanzi}}, \bibinfo {author}
  {\bibfnamefont {B.}~\bibnamefont {Deissler}}, \bibinfo {author}
  {\bibfnamefont {M.}~\bibnamefont {Inguscio}}, \bibinfo {author}
  {\bibfnamefont {G.}~\bibnamefont {Modugno}}, \bibinfo {author} {\bibfnamefont
  {M.~B.}\ \bibnamefont {Plenio}}, \ and\ \bibinfo {author} {\bibfnamefont
  {F.}~\bibnamefont {Caruso}},\ }\href@noop {} {\bibfield  {journal} {\bibinfo
  {journal} {New Journal of Physics}\ }\textbf {\bibinfo {volume} {15}},\
  \bibinfo {pages} {045007} (\bibinfo {year} {2013})}\BibitemShut {NoStop}%
\bibitem [{\citenamefont {Kraus}\ \emph {et~al.}(2012)\citenamefont {Kraus},
  \citenamefont {Lahini}, \citenamefont {Ringel}, \citenamefont {Verbin},\ and\
  \citenamefont {Zilberberg}}]{expt4}%
  \BibitemOpen
  \bibfield  {author} {\bibinfo {author} {\bibfnamefont {Y.~E.}\ \bibnamefont
  {Kraus}}, \bibinfo {author} {\bibfnamefont {Y.}~\bibnamefont {Lahini}},
  \bibinfo {author} {\bibfnamefont {Z.}~\bibnamefont {Ringel}}, \bibinfo
  {author} {\bibfnamefont {M.}~\bibnamefont {Verbin}}, \ and\ \bibinfo {author}
  {\bibfnamefont {O.}~\bibnamefont {Zilberberg}},\ }\href@noop {} {\bibfield
  {journal} {\bibinfo  {journal} {Phys. Rev. Lett.}\ }\textbf {\bibinfo
  {volume} {109}},\ \bibinfo {pages} {106402} (\bibinfo {year}
  {2012})}\BibitemShut {NoStop}%
\bibitem [{\citenamefont {Lahini}\ \emph {et~al.}(2009)\citenamefont {Lahini},
  \citenamefont {Pugatch}, \citenamefont {Pozzi}, \citenamefont {Sorel},
  \citenamefont {Morandotti}, \citenamefont {Davidson},\ and\ \citenamefont
  {Silberberg}}]{expt5}%
  \BibitemOpen
  \bibfield  {author} {\bibinfo {author} {\bibfnamefont {Y.}~\bibnamefont
  {Lahini}}, \bibinfo {author} {\bibfnamefont {R.}~\bibnamefont {Pugatch}},
  \bibinfo {author} {\bibfnamefont {F.}~\bibnamefont {Pozzi}}, \bibinfo
  {author} {\bibfnamefont {M.}~\bibnamefont {Sorel}}, \bibinfo {author}
  {\bibfnamefont {R.}~\bibnamefont {Morandotti}}, \bibinfo {author}
  {\bibfnamefont {N.}~\bibnamefont {Davidson}}, \ and\ \bibinfo {author}
  {\bibfnamefont {Y.}~\bibnamefont {Silberberg}},\ }\href@noop {} {\bibfield
  {journal} {\bibinfo  {journal} {Phys. Rev. Lett.}\ }\textbf {\bibinfo
  {volume} {103}},\ \bibinfo {pages} {013901} (\bibinfo {year}
  {2009})}\BibitemShut {NoStop}%
\bibitem [{\citenamefont {Roati}\ \emph {et~al.}(2008)\citenamefont {Roati},
  \citenamefont {\'Errico}, \citenamefont {Fallani}, \citenamefont {Fattori},
  \citenamefont {Fort}, \citenamefont {Zaccanti}, \citenamefont {Modugno},
  \citenamefont {Modugno},\ and\ \citenamefont {Inguscio}}]{expt6}%
  \BibitemOpen
  \bibfield  {author} {\bibinfo {author} {\bibfnamefont {G.}~\bibnamefont
  {Roati}}, \bibinfo {author} {\bibfnamefont {C.~D.}\ \bibnamefont {\'Errico}},
  \bibinfo {author} {\bibfnamefont {L.}~\bibnamefont {Fallani}}, \bibinfo
  {author} {\bibfnamefont {M.}~\bibnamefont {Fattori}}, \bibinfo {author}
  {\bibfnamefont {C.}~\bibnamefont {Fort}}, \bibinfo {author} {\bibfnamefont
  {M.}~\bibnamefont {Zaccanti}}, \bibinfo {author} {\bibfnamefont
  {G.}~\bibnamefont {Modugno}}, \bibinfo {author} {\bibfnamefont
  {M.}~\bibnamefont {Modugno}}, \ and\ \bibinfo {author} {\bibfnamefont
  {M.}~\bibnamefont {Inguscio}},\ }\href@noop {} {\bibfield  {journal}
  {\bibinfo  {journal} {Nature}\ }\textbf {\bibinfo {volume} {453}},\ \bibinfo
  {pages} {895} (\bibinfo {year} {2008})}\BibitemShut {NoStop}%
\bibitem [{\citenamefont {Hiramoto}\ and\ \citenamefont
  {Abe}(1988)}]{wavepacket1}%
  \BibitemOpen
  \bibfield  {author} {\bibinfo {author} {\bibfnamefont {H.}~\bibnamefont
  {Hiramoto}}\ and\ \bibinfo {author} {\bibfnamefont {S.}~\bibnamefont {Abe}},\
  }\href@noop {} {\bibfield  {journal} {\bibinfo  {journal} {Journal of the
  Physical Society of Japan}\ }\textbf {\bibinfo {volume} {57}},\ \bibinfo
  {pages} {230} (\bibinfo {year} {1988})}\BibitemShut {NoStop}%
\bibitem [{\citenamefont {Zhong}\ \emph {et~al.}(2001)\citenamefont {Zhong},
  \citenamefont {Diener}, \citenamefont {Steck}, \citenamefont {Oskay},
  \citenamefont {Raizen}, \citenamefont {Plummer}, \citenamefont {Zhang},\ and\
  \citenamefont {Niu}}]{wavepacket2}%
  \BibitemOpen
  \bibfield  {author} {\bibinfo {author} {\bibfnamefont {J.}~\bibnamefont
  {Zhong}}, \bibinfo {author} {\bibfnamefont {R.~B.}\ \bibnamefont {Diener}},
  \bibinfo {author} {\bibfnamefont {D.~A.}\ \bibnamefont {Steck}}, \bibinfo
  {author} {\bibfnamefont {W.~H.}\ \bibnamefont {Oskay}}, \bibinfo {author}
  {\bibfnamefont {M.~G.}\ \bibnamefont {Raizen}}, \bibinfo {author}
  {\bibfnamefont {E.~W.}\ \bibnamefont {Plummer}}, \bibinfo {author}
  {\bibfnamefont {Z.}~\bibnamefont {Zhang}}, \ and\ \bibinfo {author}
  {\bibfnamefont {Q.}~\bibnamefont {Niu}},\ }\href@noop {} {\bibfield
  {journal} {\bibinfo  {journal} {Phys. Rev. Lett.}\ }\textbf {\bibinfo
  {volume} {86}},\ \bibinfo {pages} {2485} (\bibinfo {year}
  {2001})}\BibitemShut {NoStop}%
\bibitem [{\citenamefont {Ng}\ and\ \citenamefont
  {Kottos}(2007)}]{wavepacket3}%
  \BibitemOpen
  \bibfield  {author} {\bibinfo {author} {\bibfnamefont {G.~S.}\ \bibnamefont
  {Ng}}\ and\ \bibinfo {author} {\bibfnamefont {T.}~\bibnamefont {Kottos}},\
  }\href@noop {} {\bibfield  {journal} {\bibinfo  {journal} {Phys. Rev. B}\
  }\textbf {\bibinfo {volume} {75}},\ \bibinfo {pages} {205120} (\bibinfo
  {year} {2007})}\BibitemShut {NoStop}%
\bibitem [{\citenamefont {Ketzmerick}\ \emph {et~al.}(1997)\citenamefont
  {Ketzmerick}, \citenamefont {Kruse}, \citenamefont {Kraut},\ and\
  \citenamefont {Geisel}}]{wavepacket4}%
  \BibitemOpen
  \bibfield  {author} {\bibinfo {author} {\bibfnamefont {R.}~\bibnamefont
  {Ketzmerick}}, \bibinfo {author} {\bibfnamefont {K.}~\bibnamefont {Kruse}},
  \bibinfo {author} {\bibfnamefont {S.}~\bibnamefont {Kraut}}, \ and\ \bibinfo
  {author} {\bibfnamefont {T.}~\bibnamefont {Geisel}},\ }\href@noop {}
  {\bibfield  {journal} {\bibinfo  {journal} {Phys. Rev. Lett.}\ }\textbf
  {\bibinfo {volume} {79}},\ \bibinfo {pages} {1959} (\bibinfo {year}
  {1997})}\BibitemShut {NoStop}%
\bibitem [{\citenamefont {Everest}\ \emph {et~al.}(2017)\citenamefont
  {Everest}, \citenamefont {Lesanovsky}, \citenamefont {Garrahan},\ and\
  \citenamefont {Levi}}]{MBL_open_1}%
  \BibitemOpen
  \bibfield  {author} {\bibinfo {author} {\bibfnamefont {B.}~\bibnamefont
  {Everest}}, \bibinfo {author} {\bibfnamefont {I.}~\bibnamefont {Lesanovsky}},
  \bibinfo {author} {\bibfnamefont {J.~P.}\ \bibnamefont {Garrahan}}, \ and\
  \bibinfo {author} {\bibfnamefont {E.}~\bibnamefont {Levi}},\ }\href {\doibase
  10.1103/PhysRevB.95.024310} {\bibfield  {journal} {\bibinfo  {journal} {Phys.
  Rev. B}\ }\textbf {\bibinfo {volume} {95}},\ \bibinfo {pages} {024310}
  (\bibinfo {year} {2017})}\BibitemShut {NoStop}%
\bibitem [{\citenamefont {Levi}\ \emph {et~al.}(2016)\citenamefont {Levi},
  \citenamefont {Heyl}, \citenamefont {Lesanovsky},\ and\ \citenamefont
  {Garrahan}}]{MBL_open_2}%
  \BibitemOpen
  \bibfield  {author} {\bibinfo {author} {\bibfnamefont {E.}~\bibnamefont
  {Levi}}, \bibinfo {author} {\bibfnamefont {M.}~\bibnamefont {Heyl}}, \bibinfo
  {author} {\bibfnamefont {I.}~\bibnamefont {Lesanovsky}}, \ and\ \bibinfo
  {author} {\bibfnamefont {J.~P.}\ \bibnamefont {Garrahan}},\ }\href {\doibase
  10.1103/PhysRevLett.116.237203} {\bibfield  {journal} {\bibinfo  {journal}
  {Phys. Rev. Lett.}\ }\textbf {\bibinfo {volume} {116}},\ \bibinfo {pages}
  {237203} (\bibinfo {year} {2016})}\BibitemShut {NoStop}%
\bibitem [{\citenamefont {Fischer}\ \emph {et~al.}(2016)\citenamefont
  {Fischer}, \citenamefont {Maksymenko},\ and\ \citenamefont
  {Altman}}]{MBL_open_3}%
  \BibitemOpen
  \bibfield  {author} {\bibinfo {author} {\bibfnamefont {M.~H.}\ \bibnamefont
  {Fischer}}, \bibinfo {author} {\bibfnamefont {M.}~\bibnamefont {Maksymenko}},
  \ and\ \bibinfo {author} {\bibfnamefont {E.}~\bibnamefont {Altman}},\ }\href
  {\doibase 10.1103/PhysRevLett.116.160401} {\bibfield  {journal} {\bibinfo
  {journal} {Phys. Rev. Lett.}\ }\textbf {\bibinfo {volume} {116}},\ \bibinfo
  {pages} {160401} (\bibinfo {year} {2016})}\BibitemShut {NoStop}%
\bibitem [{\citenamefont {Nandkishore}\ and\ \citenamefont
  {Gopalakrishnan}(2016)}]{MBL_open_4}%
  \BibitemOpen
  \bibfield  {author} {\bibinfo {author} {\bibfnamefont {R.}~\bibnamefont
  {Nandkishore}}\ and\ \bibinfo {author} {\bibfnamefont {S.}~\bibnamefont
  {Gopalakrishnan}},\ }\href {\doibase 10.1002/andp.201600181} {\bibfield
  {journal} {\bibinfo  {journal} {Annalen der Physik}\ } (\bibinfo {year}
  {2016}),\ 10.1002/andp.201600181}\BibitemShut {NoStop}%
\bibitem [{\citenamefont {Banerjee}\ and\ \citenamefont
  {Altman}(2016)}]{MBL_open_5}%
  \BibitemOpen
  \bibfield  {author} {\bibinfo {author} {\bibfnamefont {S.}~\bibnamefont
  {Banerjee}}\ and\ \bibinfo {author} {\bibfnamefont {E.}~\bibnamefont
  {Altman}},\ }\href {\doibase 10.1103/PhysRevLett.116.116601} {\bibfield
  {journal} {\bibinfo  {journal} {Phys. Rev. Lett.}\ }\textbf {\bibinfo
  {volume} {116}},\ \bibinfo {pages} {116601} (\bibinfo {year}
  {2016})}\BibitemShut {NoStop}%
\bibitem [{\citenamefont {Johri}\ \emph {et~al.}(2015)\citenamefont {Johri},
  \citenamefont {Nandkishore},\ and\ \citenamefont {Bhatt}}]{MBL_open_6}%
  \BibitemOpen
  \bibfield  {author} {\bibinfo {author} {\bibfnamefont {S.}~\bibnamefont
  {Johri}}, \bibinfo {author} {\bibfnamefont {R.}~\bibnamefont {Nandkishore}},
  \ and\ \bibinfo {author} {\bibfnamefont {R.~N.}\ \bibnamefont {Bhatt}},\
  }\href {\doibase 10.1103/PhysRevLett.114.117401} {\bibfield  {journal}
  {\bibinfo  {journal} {Phys. Rev. Lett.}\ }\textbf {\bibinfo {volume} {114}},\
  \bibinfo {pages} {117401} (\bibinfo {year} {2015})}\BibitemShut {NoStop}%
\bibitem [{\citenamefont {Huse}\ \emph {et~al.}(2015)\citenamefont {Huse},
  \citenamefont {Nandkishore}, \citenamefont {Pietracaprina}, \citenamefont
  {Ros},\ and\ \citenamefont {Scardicchio}}]{MBL_open_7}%
  \BibitemOpen
  \bibfield  {author} {\bibinfo {author} {\bibfnamefont {D.~A.}\ \bibnamefont
  {Huse}}, \bibinfo {author} {\bibfnamefont {R.}~\bibnamefont {Nandkishore}},
  \bibinfo {author} {\bibfnamefont {F.}~\bibnamefont {Pietracaprina}}, \bibinfo
  {author} {\bibfnamefont {V.}~\bibnamefont {Ros}}, \ and\ \bibinfo {author}
  {\bibfnamefont {A.}~\bibnamefont {Scardicchio}},\ }\href {\doibase
  10.1103/PhysRevB.92.014203} {\bibfield  {journal} {\bibinfo  {journal} {Phys.
  Rev. B}\ }\textbf {\bibinfo {volume} {92}},\ \bibinfo {pages} {014203}
  (\bibinfo {year} {2015})}\BibitemShut {NoStop}%
\bibitem [{\citenamefont {Nandkishore}(2015)}]{MBL_open_8}%
  \BibitemOpen
  \bibfield  {author} {\bibinfo {author} {\bibfnamefont {R.}~\bibnamefont
  {Nandkishore}},\ }\href {\doibase 10.1103/PhysRevB.92.245141} {\bibfield
  {journal} {\bibinfo  {journal} {Phys. Rev. B}\ }\textbf {\bibinfo {volume}
  {92}},\ \bibinfo {pages} {245141} (\bibinfo {year} {2015})}\BibitemShut
  {NoStop}%
\bibitem [{\citenamefont {Nandkishore}\ \emph {et~al.}(2014)\citenamefont
  {Nandkishore}, \citenamefont {Gopalakrishnan},\ and\ \citenamefont
  {Huse}}]{MBL_open_9}%
  \BibitemOpen
  \bibfield  {author} {\bibinfo {author} {\bibfnamefont {R.}~\bibnamefont
  {Nandkishore}}, \bibinfo {author} {\bibfnamefont {S.}~\bibnamefont
  {Gopalakrishnan}}, \ and\ \bibinfo {author} {\bibfnamefont {D.~A.}\
  \bibnamefont {Huse}},\ }\href {\doibase 10.1103/PhysRevB.90.064203}
  {\bibfield  {journal} {\bibinfo  {journal} {Phys. Rev. B}\ }\textbf {\bibinfo
  {volume} {90}},\ \bibinfo {pages} {064203} (\bibinfo {year}
  {2014})}\BibitemShut {NoStop}%
\bibitem [{\citenamefont {Gopalakrishnan}\ and\ \citenamefont
  {Nandkishore}(2014)}]{MBL_open_10}%
  \BibitemOpen
  \bibfield  {author} {\bibinfo {author} {\bibfnamefont {S.}~\bibnamefont
  {Gopalakrishnan}}\ and\ \bibinfo {author} {\bibfnamefont {R.}~\bibnamefont
  {Nandkishore}},\ }\href {\doibase 10.1103/PhysRevB.90.224203} {\bibfield
  {journal} {\bibinfo  {journal} {Phys. Rev. B}\ }\textbf {\bibinfo {volume}
  {90}},\ \bibinfo {pages} {224203} (\bibinfo {year} {2014})}\BibitemShut
  {NoStop}%
\bibitem [{\citenamefont {Wang}\ \emph {et~al.}(2016)\citenamefont {Wang},
  \citenamefont {Zhang},\ and\ \citenamefont {Zhao}}]{nongauss_diff0}%
  \BibitemOpen
  \bibfield  {author} {\bibinfo {author} {\bibfnamefont {J.}~\bibnamefont
  {Wang}}, \bibinfo {author} {\bibfnamefont {Y.}~\bibnamefont {Zhang}}, \ and\
  \bibinfo {author} {\bibfnamefont {H.}~\bibnamefont {Zhao}},\ }\href {\doibase
  10.1103/PhysRevE.93.032144} {\bibfield  {journal} {\bibinfo  {journal} {Phys.
  Rev. E}\ }\textbf {\bibinfo {volume} {93}},\ \bibinfo {pages} {032144}
  (\bibinfo {year} {2016})}\BibitemShut {NoStop}%
\bibitem [{\citenamefont {Forte}\ \emph {et~al.}(2014)\citenamefont {Forte},
  \citenamefont {Cecconi},\ and\ \citenamefont {Vulpiani}}]{nongauss_diff1}%
  \BibitemOpen
  \bibfield  {author} {\bibinfo {author} {\bibfnamefont {G.}~\bibnamefont
  {Forte}}, \bibinfo {author} {\bibfnamefont {F.}~\bibnamefont {Cecconi}}, \
  and\ \bibinfo {author} {\bibfnamefont {A.}~\bibnamefont {Vulpiani}},\ }\href
  {\doibase 10.1140/epjb/e2014-40956-0} {\bibfield  {journal} {\bibinfo
  {journal} {The European Physical Journal B}\ }\textbf {\bibinfo {volume}
  {87}},\ \bibinfo {pages} {102} (\bibinfo {year} {2014})}\BibitemShut
  {NoStop}%
\bibitem [{\citenamefont {Chubynsky}\ and\ \citenamefont
  {Slater}(2014)}]{nongauss_diff2}%
  \BibitemOpen
  \bibfield  {author} {\bibinfo {author} {\bibfnamefont {M.~V.}\ \bibnamefont
  {Chubynsky}}\ and\ \bibinfo {author} {\bibfnamefont {G.~W.}\ \bibnamefont
  {Slater}},\ }\href {\doibase 10.1103/PhysRevLett.113.098302} {\bibfield
  {journal} {\bibinfo  {journal} {Phys. Rev. Lett.}\ }\textbf {\bibinfo
  {volume} {113}},\ \bibinfo {pages} {098302} (\bibinfo {year}
  {2014})}\BibitemShut {NoStop}%
\bibitem [{\citenamefont {Kim}\ \emph {et~al.}(2013)\citenamefont {Kim},
  \citenamefont {Kim},\ and\ \citenamefont {Sung}}]{nongauss_diff3}%
  \BibitemOpen
  \bibfield  {author} {\bibinfo {author} {\bibfnamefont {J.}~\bibnamefont
  {Kim}}, \bibinfo {author} {\bibfnamefont {C.}~\bibnamefont {Kim}}, \ and\
  \bibinfo {author} {\bibfnamefont {B.~J.}\ \bibnamefont {Sung}},\ }\href
  {\doibase 10.1103/PhysRevLett.110.047801} {\bibfield  {journal} {\bibinfo
  {journal} {Phys. Rev. Lett.}\ }\textbf {\bibinfo {volume} {110}},\ \bibinfo
  {pages} {047801} (\bibinfo {year} {2013})}\BibitemShut {NoStop}%
\bibitem [{\citenamefont {{Wang}}\ \emph {et~al.}(2012)\citenamefont {{Wang}},
  \citenamefont {{Kuo}}, \citenamefont {{Bae}},\ and\ \citenamefont
  {{Granick}}}]{nongauss_diff4}%
  \BibitemOpen
  \bibfield  {author} {\bibinfo {author} {\bibfnamefont {B.}~\bibnamefont
  {{Wang}}}, \bibinfo {author} {\bibfnamefont {J.}~\bibnamefont {{Kuo}}},
  \bibinfo {author} {\bibfnamefont {S.~C.}\ \bibnamefont {{Bae}}}, \ and\
  \bibinfo {author} {\bibfnamefont {S.}~\bibnamefont {{Granick}}},\ }\href
  {\doibase 10.1038/nmat3308} {\bibfield  {journal} {\bibinfo  {journal}
  {Nature Materials}\ }\textbf {\bibinfo {volume} {11}},\ \bibinfo {pages}
  {481} (\bibinfo {year} {2012})}\BibitemShut {NoStop}%
\bibitem [{\citenamefont {Wang}\ \emph {et~al.}(2009)\citenamefont {Wang},
  \citenamefont {Anthony}, \citenamefont {Bae},\ and\ \citenamefont
  {Granick}}]{nongauss_diff5}%
  \BibitemOpen
  \bibfield  {author} {\bibinfo {author} {\bibfnamefont {B.}~\bibnamefont
  {Wang}}, \bibinfo {author} {\bibfnamefont {S.~M.}\ \bibnamefont {Anthony}},
  \bibinfo {author} {\bibfnamefont {S.~C.}\ \bibnamefont {Bae}}, \ and\
  \bibinfo {author} {\bibfnamefont {S.}~\bibnamefont {Granick}},\ }\href
  {\doibase 10.1073/pnas.0903554106} {\bibfield  {journal} {\bibinfo  {journal}
  {Proceedings of the National Academy of Sciences}\ }\textbf {\bibinfo
  {volume} {106}},\ \bibinfo {pages} {15160} (\bibinfo {year}
  {2009})}\BibitemShut {NoStop}%
\bibitem [{\citenamefont {Andersen}\ \emph {et~al.}(2000)\citenamefont
  {Andersen}, \citenamefont {Castiglione}, \citenamefont {Mazzino},\ and\
  \citenamefont {Vulpiani}}]{nongauss_diff6}%
  \BibitemOpen
  \bibfield  {author} {\bibinfo {author} {\bibfnamefont {K.~H.}\ \bibnamefont
  {Andersen}}, \bibinfo {author} {\bibfnamefont {P.}~\bibnamefont
  {Castiglione}}, \bibinfo {author} {\bibfnamefont {A.}~\bibnamefont
  {Mazzino}}, \ and\ \bibinfo {author} {\bibfnamefont {A.}~\bibnamefont
  {Vulpiani}},\ }\href {\doibase 10.1007/s100510070032} {\bibfield  {journal}
  {\bibinfo  {journal} {Eur. Phys. J. B}\ }\textbf {\bibinfo {volume} {18}},\
  \bibinfo {pages} {447} (\bibinfo {year} {2000})}\BibitemShut {NoStop}%
\bibitem [{\citenamefont {Kong}\ and\ \citenamefont
  {Cohen}(1989)}]{nongauss_diff7}%
  \BibitemOpen
  \bibfield  {author} {\bibinfo {author} {\bibfnamefont {X.~P.}\ \bibnamefont
  {Kong}}\ and\ \bibinfo {author} {\bibfnamefont {E.~G.~D.}\ \bibnamefont
  {Cohen}},\ }\href@noop {} {\bibfield  {journal} {\bibinfo  {journal} {Phys.
  Rev. B}\ }\textbf {\bibinfo {volume} {40}},\ \bibinfo {pages} {4838}
  (\bibinfo {year} {1989})}\BibitemShut {NoStop}%
\bibitem [{\citenamefont {Purkayastha}\ \emph {et~al.}(2016)\citenamefont
  {Purkayastha}, \citenamefont {Dhar},\ and\ \citenamefont {Kulkarni}}]{ap1}%
  \BibitemOpen
  \bibfield  {author} {\bibinfo {author} {\bibfnamefont {A.}~\bibnamefont
  {Purkayastha}}, \bibinfo {author} {\bibfnamefont {A.}~\bibnamefont {Dhar}}, \
  and\ \bibinfo {author} {\bibfnamefont {M.}~\bibnamefont {Kulkarni}},\
  }\href@noop {} {\bibfield  {journal} {\bibinfo  {journal} {Phys. Rev. A}\
  }\textbf {\bibinfo {volume} {93}},\ \bibinfo {pages} {062114} (\bibinfo
  {year} {2016})}\BibitemShut {NoStop}%
\bibitem [{\citenamefont {Purkayastha}(2017)}]{Archak2017}%
  \BibitemOpen
  \bibfield  {author} {\bibinfo {author} {\bibfnamefont {A.}~\bibnamefont
  {Purkayastha}},\ }\href@noop {} {\bibfield  {journal} {\bibinfo  {journal}
  {arXiv:1712.01068}\ } (\bibinfo {year} {2017})}\BibitemShut {NoStop}%
\bibitem [{\citenamefont {De~Roeck}\ \emph {et~al.}(2017)\citenamefont
  {De~Roeck}, \citenamefont {Dhar}, \citenamefont {Huveneers},\ and\
  \citenamefont {Sch{\"u}tz}}]{dhar2017}%
  \BibitemOpen
  \bibfield  {author} {\bibinfo {author} {\bibfnamefont {W.}~\bibnamefont
  {De~Roeck}}, \bibinfo {author} {\bibfnamefont {A.}~\bibnamefont {Dhar}},
  \bibinfo {author} {\bibfnamefont {F.}~\bibnamefont {Huveneers}}, \ and\
  \bibinfo {author} {\bibfnamefont {M.}~\bibnamefont {Sch{\"u}tz}},\ }\href
  {\doibase 10.1007/s10955-017-1769-z} {\bibfield  {journal} {\bibinfo
  {journal} {Journal of Statistical Physics}\ }\textbf {\bibinfo {volume}
  {167}},\ \bibinfo {pages} {1143} (\bibinfo {year} {2017})}\BibitemShut
  {NoStop}%
\bibitem [{\citenamefont {Amir}\ \emph {et~al.}(2018)\citenamefont {Amir},
  \citenamefont {Oreg},\ and\ \citenamefont {Imry}}]{amir2018}%
  \BibitemOpen
  \bibfield  {author} {\bibinfo {author} {\bibfnamefont {A.}~\bibnamefont
  {Amir}}, \bibinfo {author} {\bibfnamefont {Y.}~\bibnamefont {Oreg}}, \ and\
  \bibinfo {author} {\bibfnamefont {Y.}~\bibnamefont {Imry}},\ }\href@noop {}
  {\bibfield  {journal} {\bibinfo  {journal} {arXiv:1801.09707}\ } (\bibinfo
  {year} {2018})}\BibitemShut {NoStop}%
\bibitem [{\citenamefont {Purkayastha}(2018)}]{archak2018}%
  \BibitemOpen
  \bibfield  {author} {\bibinfo {author} {\bibfnamefont {A.}~\bibnamefont
  {Purkayastha}},\ }\href@noop {} {\bibfield  {journal} {\bibinfo  {journal}
  {in preparation}\ } (\bibinfo {year} {2018})}\BibitemShut {NoStop}%
\bibitem [{\citenamefont {Monthus}(2017)}]{monthus}%
  \BibitemOpen
  \bibfield  {author} {\bibinfo {author} {\bibfnamefont {C.}~\bibnamefont
  {Monthus}},\ }\href {http://stacks.iop.org/1742-5468/2017/i=4/a=043303}
  {\bibfield  {journal} {\bibinfo  {journal} {Journal of Statistical Mechanics:
  Theory and Experiment}\ }\textbf {\bibinfo {volume} {2017}},\ \bibinfo
  {pages} {043303} (\bibinfo {year} {2017})}\BibitemShut {NoStop}%
\bibitem [{\citenamefont {Varma}\ \emph {et~al.}(2017)\citenamefont {Varma},
  \citenamefont {de~Mulatier},\ and\ \citenamefont {\ifmmode \check{Z}\else
  \v{Z}\fi{}nidari\ifmmode~\check{c}\else \v{c}\fi{}}}]{vkv}%
  \BibitemOpen
  \bibfield  {author} {\bibinfo {author} {\bibfnamefont {V.~K.}\ \bibnamefont
  {Varma}}, \bibinfo {author} {\bibfnamefont {C.}~\bibnamefont {de~Mulatier}},
  \ and\ \bibinfo {author} {\bibfnamefont {M.}~\bibnamefont {\ifmmode
  \check{Z}\else \v{Z}\fi{}nidari\ifmmode~\check{c}\else \v{c}\fi{}}},\ }\href
  {\doibase 10.1103/PhysRevE.96.032130} {\bibfield  {journal} {\bibinfo
  {journal} {Phys. Rev. E}\ }\textbf {\bibinfo {volume} {96}},\ \bibinfo
  {pages} {032130} (\bibinfo {year} {2017})}\BibitemShut {NoStop}%
\bibitem [{\citenamefont {\ifmmode \check{Z}\else
  \v{Z}\fi{}nidari\ifmmode~\check{c}\else \v{c}\fi{}}\ \emph
  {et~al.}(2016)\citenamefont {\ifmmode \check{Z}\else
  \v{Z}\fi{}nidari\ifmmode~\check{c}\else \v{c}\fi{}}, \citenamefont
  {Scardicchio},\ and\ \citenamefont {Varma}}]{Znidaric_sub_diff}%
  \BibitemOpen
  \bibfield  {author} {\bibinfo {author} {\bibfnamefont {M.}~\bibnamefont
  {\ifmmode \check{Z}\else \v{Z}\fi{}nidari\ifmmode~\check{c}\else
  \v{c}\fi{}}}, \bibinfo {author} {\bibfnamefont {A.}~\bibnamefont
  {Scardicchio}}, \ and\ \bibinfo {author} {\bibfnamefont {V.~K.}\ \bibnamefont
  {Varma}},\ }\href {\doibase 10.1103/PhysRevLett.117.040601} {\bibfield
  {journal} {\bibinfo  {journal} {Phys. Rev. Lett.}\ }\textbf {\bibinfo
  {volume} {117}},\ \bibinfo {pages} {040601} (\bibinfo {year}
  {2016})}\BibitemShut {NoStop}%
\end{thebibliography}%

\end{document}